\documentclass{aa}  
\DeclareUnicodeCharacter{2212}{-}

\usepackage{graphicx}
\usepackage{txfonts}
\usepackage{amsmath}
\usepackage[labelfont=bf, labelsep=period]{caption}
\usepackage{subcaption}
\usepackage{sidecap}
\usepackage{titlesec}
\usepackage{multirow}
\usepackage{xcolor}
\usepackage{natbib}
\usepackage{lineno}

\usepackage{hyperref}
\usepackage{cleveref}

\hypersetup{
    colorlinks=true,        
    linkcolor=blue,         
    citecolor=blue,         
    filecolor=magenta,      
    urlcolor=blue         
} 

\setcitestyle{authoryear}
\usepackage{orcidlink}

\usepackage{xcolor}

\definecolor{refblue}{RGB}{0,51,153}
\definecolor{refred}{RGB}{180,0,0}
\definecolor{refgreen}{RGB}{0,102,51}

\title{The caustic method applied to The Three Hundred: prospects for upcoming CATARSIS and other surveys}
\titlerunning{The Caustic Method Applied to The300}

\author{
B. Callejas-Córdoba\inst{1,2}\orcidlink{0009-0002-6256-8294}
\and
P. Sánchez-Blázquez\inst{1,2}\orcidlink{0000-0003-0651-0098}
\and
A. Gil de Paz\inst{1,2}\orcidlink{0000-0001-6150-2854}
\and
A. Knebe\inst{3,4,5}\orcidlink{0000-0003-4066-8307}
\and
W. Cui\inst{3,4,6}\orcidlink{0000-0002-2113-4863}
\and
C. Catalán-Torrecilla\inst{1,2}\orcidlink{0000-0002-8067-0164}
\and R. Dave \inst{6}\orcidlink{0000-0003-2842-9434}
}

\institute{\itshape
$^{1}$ Departamento de Física de la Tierra y Astrofísica, Facultad de Ciencias Físicas, Universidad Complutense de Madrid, Plaza de las Ciencias 1, Madrid, Spain\\
$^{2}$ Instituto de Física de Partículas y del Cosmos (IPARCOS-UCM), Facultad de Ciencias Físicas, Universidad Complutense de Madrid, Plaza de las Ciencias 1, 28040 Madrid, Spain\\
$^{3}$ Departamento de F\'isica Te\'orica, Facultad de Ciencias, Universidad Aut\'onoma de Madrid, 28049 Madrid, Spain\\
$^{4}$ Centro de Investigaci\'on Avanzada en F\'isica Fundamental (CIAFF), Facultad de Ciencias, Universidad Aut\'onoma de Madrid, 28049 Madrid, Spain\\
$^{5}$ International center for Radio Astronomy Research, University of Western Australia, 35 Stirling Highway, Crawley, Western Australia 6009, Australia\\
$^{6}$ Institute for Astronomy, University of Edinburgh, Royal Observatory, Blackford Hill, Edinburgh, EH9 3HJ, UK
}
\authorrunning{Callejas et al.}

\date{}\href{}{}

\abstract{We investigate the expected uncertainties in recovering galaxy cluster mass profiles from upcoming spectroscopic survey data using The Three Hundred Project. Using the caustic technique, which leverages galaxy positions and line-of-sight velocities, we assess the systematic errors introduced by assumptions regarding velocity anisotropy and demonstrate how an iterative correction method can minimize these errors. We also assess the impact of survey magnitude limits on cluster mass estimates, highlighting potential biases across different observational strategies We focus the analysis on our own CATARSIS survey, which aims at obtaining redshift measurements for all galaxies with magnitudes $m_{AB, r} < 22$ within 2$\times$$R_{\mathrm{200,c}}$ of 16 galaxy clusters with redshifts  $0.14 < z < 0.27$ using the future 8\,arcmin$^2$ field-of-view TARSIS integral-field spectrograph of the Calar Alto 3.5-m telescope. Such data will enable us to mitigate systematic errors in the determination of density profiles. CATARSIS aims at enhancing the precision of mass profile estimates by deepening our understanding of the dynamical states and physical characteristics of galaxy clusters.}


\begin{document}

\maketitle
\nolinenumbers

\vspace{1em}  

\section{Introduction}
    The observational determination of the masses of galaxy clusters is of central importance to our understanding of the growth of structure in the Universe and the use of clusters as cosmological probes. The distribution of mass within galaxy clusters, that is, the mass profiles, has allowed not only to elucidate the nature of dark matter (i.e$.$ warm vs$.$ cold particles; \citealt{blumenthal_formation_1984}) but also to distinguish between possible alternatives to the existence of this component, as modified gravity theories (see, e.g., \citealt{milgrom_mond--theoretical_2002}; \citealt{mcgaugh_tale_2015}). The determination of mass profiles in clusters is also an essential reference point for studies of the astrophysical processes shaping the properties of the baryons in clusters, both the intra-cluster medium (ICM) and the member galaxies.
    
    Galaxy cluster masses can be derived using a variety of observational techniques. Each approach has specific assumptions and systematic limitations. Techniques based on dynamical equilibrium, such as X-ray, Sunyaev--Zel'dovich (SZ) determinations \citep{sarazin_1988}, and Jeans analyses (\citealt{the_and_white_1986}; \citealt{merritt_distribution_1987}; \citealt{Binney_and_Tremaine_2008}; \citealt{MamonLokas_2003}; \citealt{Lokas2006}; \citealt{MamonBoue2010}), can become unreliable even within $R_{\mathrm{200,c}}$ when probing the cluster outskirts, where the assumption of equilibrium starts to break down and clusters continue to accrete galaxies and dark matter (\citealt{ludlow_dynamical_2012}; \citealt{bakels_pre-processing_2021}). 
    In contrast, weak gravitational lensing (\citealt{bartelmann_2010}; \citealt{umetsu_weak-lensing_2020}) and the caustic technique (\citealt{diaferio_infall_1997}; \citealt{diaferio_mass_1999}; \citealt{serra_measuring_2010}) provide reliable mass estimates at large radii without requiring equilibrium assumptions. In addition, several techniques based on galaxy tracers have been developed, including mass--richness relations, optical probability distribution functions (OPDF; \citealt{Li_2021}), intra-cluster light (ICL; \citealt{Contreras_2024}), and machine-learning methods (\citealt{DeAndres_2024}), which offer complementary approaches to estimate cluster masses from optical data.
    
    X-ray observations can suffer from biases due to non-thermal pressure sources such as gas accretion, AGN feedback, substructures, turbulence, and cosmic rays, leading to mass underestimations of 10–30\% (\citealt{rasia_lensing_2012}; \citealt{nelson_weighing_2014}). Weak lensing is affected by the presence of interlopers and projection effects, due to the triaxial shape of dark matter halos, which can lead to overestimated masses by up to 40\% (\citealt{corless_new_2009}; \citealt{becker_accuracy_2011}; \citealt{feroz_weak_2012}).
    For completeness, it is worth noting that other mass proxies, such as the integrated SZ signal, can also suffer from limitations, particularly in the cluster outskirts where the signal-to-background ratio becomes low, making SZ-based mass estimates less reliable (\citealt{yoon_outskirts_2022}).
    
    The caustic method provides a dynamical way to estimate cluster masses from the galaxy velocity field by identifying the escape velocity profile in projected phase space. 
    The caustic technique, using galaxy redshift data, offers an alternative by mapping the positions and velocities of cluster members, revealing a characteristic ``trumpet-shaped'' profile that traces the mass distribution without assuming equilibrium \citep{diaferio_infall_1997,diaferio_mass_1999,serra_measuring_2010,gifford_systematic_2013}. However, the method depends on a calibration factor, $F_{\beta}(r)$, that encodes the line-of-sight projection of the velocity anisotropy profile $\beta(r)$. Adopting a constant or a single average anisotropy for all clusters can introduce systematic biases: simulation studies and empirical tests report overestimates of the enclosed mass in the inner regions of order tens of percent (commonly $\sim 30$--$70\%$ at small radii in some tests) and a typical per-cluster scatter of $\sim 30$--$50\%$ depending on spectroscopic sampling and projection effects \citep{diaferio_mass_1999,serra_measuring_2010,munari_effects_2013,pizzardo_illustristng_2023}. The bias decreases for well-sampled clusters---with $N_{\rm gal}\gtrsim150$ the average bias can be $\lesssim 5\%$ while for sparse samples (tens of members) both the bias and scatter rise sharply \citep{gifford_systematic_2013} . To mitigate the bias introduced by adopting a constant anisotropy profile, previous works have calibrated an average $F_{\beta}(r)$ from numerical simulations \citep{serra_measuring_2010, gifford_systematic_2013}. Even after this calibration, an additional uncertainty of order tens of percent must be included to account for residual anisotropy assumptions and projection effects.

    Comparative studies have revealed discrepancies between methods.  \cite{maughan_hydrostatic_2016} found that X-ray masses were ~20\% higher than caustic masses, potentially due to assumptions about the caustic filling factor (\citealt{diaferio_mass_1999}; \citealt{serra_measuring_2010}; \citealt{gifford_systematic_2013}). \citet{foex_comparison_2017} studied 10 galaxy clusters and found that Jeans, caustic, and virial mass estimates to be 20\%, 30\%, and 50\% higher than X-ray masses, respectively, with differences diminishing when accounted for substructures. Weak lensing provides direct mass estimates that do not rely on dynamical or hydrostatic equilibrium, making it a critical benchmark for calibrating other techniques. Systematic differences of order $\sim$10-30\% have been reported between weak-lensing and X-ray or dynamical mass estimates, particularly at cluster outskirts, i.e. beyond the virial radius where equilibrium assumptions break down \citep{hoekstra_how_2003,hoekstra_et_al_2011,umetsu_cluster_2011,umetsu_model-free_2013,umetsu_weak-lensing_2020}. Weak lensing and the caustic technique offer complementary approaches for studying cluster outskirts, with an agreement ranging $\sim$20-30\% \citep{diaferio_caustic_2005,geller_measuring_2013}. 

    Large spectroscopic surveys, such as the Cluster Infall Regions in the Sloan Digital Sky Survey (CIRS) and the Hectospec Cluster Survey (HeCS), have characterized the mass profiles of $\sim$130 galaxy clusters \citep{rines_cirs_2006,rines_measuring_2013}. \citet{pizzardo_mass_2021} using the caustic method. Upcoming wide-field surveys with new spectroscopic instruments \citep{dalton_et_al_2012,Tamura_2016,mcclintock_dark_2019} and multi-wavelength facilities will provide cluster samples selected through scaling relations (e.g. SZ, X-ray luminosity). These will require calibration of systematic effects using refined mass measurements from weak lensing and caustics, in order to achieve robust cosmological constraints.

    Cosmological simulations play a crucial role in understanding the formation and evolution of galaxy clusters. Obtaining representative galaxy clusters in hydrodynamical simulations is challenging because, within typical cosmological volumes, the number of massive clusters is small. Therefore, dedicated zoom-in simulations are often performed, selecting halos from dark-matter-only runs and resimulating the cluster regions with higher resolution and baryonic physics (e.g., TNG Cluster, \sc \large The Three Hundred \normalsize \rm project) \citep{Nelson_2024, cui_three_2018}. Current simulations allow the study of galaxy clusters under controlled conditions, enabling an assessment of biases in mass determination introduced by observational limitations, such as the number of galaxies available, sample selection, or spectroscopic measurement errors \citep{diaferio_mass_1999, serra_measuring_2010, gifford_systematic_2013, deAndres_2022, gianfagna_hydrostatic_2022, Gianfagna_2023, Ferragamo_2023, Iqbal_2025}.

    In this paper, we use the numerical simulations of The Three Hundred Project to present a systematic study of the accuracy of galaxy cluster density profiles recovered with the caustic technique. While similar analyses have been performed in the past, our work provides a more detailed assessment of the impact of sample selection, explicitly accounting for the observational constraints of the CATARSIS survey and enabling direct comparisons with other spectroscopic surveys.  The galaxy population is resolved down to stellar mass of $M_\star \gtrsim 10^8 M_\odot$ in the simulations, which is sufficient to study the expected performance of mass estimators; future surveys will probe even lower-mass galaxies. In Sections~\ref{catarsis} and \ref{The300}, we describe the main characteristics of the CATARSIS survey and the \sc \large The Three Hundred \normalsize \rm simulations. Sections~\ref{Caustic Technique} and \ref{Iterative Method} summarize the caustic method and the modifications introduced in this work, while Section~\ref{context of catarsis} applies this method to the expected data that will be collected by CATARSIS. Section~\ref{other_surveys} summarizes the prospects for other ongoing and future spectroscopic surveys focused on galaxy clusters. Finally, we summarize our findings in Section~\ref{conclusions}. Throughout the paper, we use the Planck cosmological parameters \citep{planck_collaboration_planck_2016}, with $\Omega_{m0} = 0.3089$, $\Omega_{\Lambda 0} = 0.6911$, and $H_{0} = 67.74 \ \rm km \ s^{-1} \ Mpc^{-1}$.

\section{CATARSIS} \label{catarsis}

    The Calar Alto Tetra-Armed Super-IFU Spectrograph Survey (CATARSIS; \citealt{Armando_2024}) is a new legacy program to be carried out with the 3.5-m telescope at the Calar Alto Observatory (Spain), using the dedicated TARSIS instrument. TARSIS is an integral-field spectrograph (IFS) with a $2.8\times2.8$ arcmin$^2$ field of view and 2 arcsec spatial resolution, consisting of four spectrographs that together cover the wavelength range 320–810 nm. CATARSIS is designed as a blind spectroscopic survey of galaxy clusters in the redshift range $0.14 < z < 0.27$, sampling systems with a broad range of masses and dynamical states. The planned deep exposures will provide an unprecedented number of spectroscopic members, yielding $\sim$500–1000 redshifts for clusters of mass $(5$–$10)\times10^{14} M_{\odot}$ at the targeted redshifts. With CATARSIS we can observe galaxies down to a limiting magnitude of $m_{r,\mathrm{lim}}\sim21$ mag without any photometric pre-selection, as the survey is carried out using blind wide-field IFS observations. This approach also enables the recovery of redshifts for dust-reddened cluster members, which are often misclassified as background galaxies in photometrically selected samples. The use of an IFS will therefore significantly reduce the biases introduced by other surveys that rely on the brightest galaxies to confirm cluster membership and are affected by fiber collisions or slit-placement constraints.

    CATARSIS aims to deliver major contributions to our understanding of galaxy evolution in dynamically evolving environments. Its scientific goals include probing the nature of dark matter and dark energy, testing cosmological models, and investigating the role of the environment in shaping galaxy properties.

\section{The Three Hundred} \label{The300}

    \sc \large The Three Hundred \normalsize \rm project \footnote{\url{https://www.nottingham.ac.uk/astronomy/The300/index.php}} (\citealt{cui_three_2018}; \textsc{The300} from now on) consists of cluster-scale zoom simulations based on a comprehensive sample of the 324 most massive dark matter halos ($M_{\rm vir} \gtrsim 8 \times 10^{14} \, h^{-1} \, \rm M_{\odot}$) selected from the MultiDark simulation (MDPL2; \citealt{klypin_multidark_2016}). MDPL2 assumes cosmological parameters from Planck \citep{planck_collaboration_planck_2016} and has a periodic cube of comoving length $1 \, h^{-1} \rm Gpc$.

    For each cluster halo, a cubic region of $15\,h^{-1}\,\rm Mpc$ comoving size was resimulated at higher resolution using different galaxy formation approaches, including hydrodynamical codes with varying sub-grid physics prescriptions—GADGET-MUSIC \citep{sembolini_music_2013}, GADGET-X \citep{rasia_cool_2015, Beck2016}, and more recently GIZMO-{\sc{Simba}} \citep{cui_span_2022}—as well as semi-analytical models (SAMs). The high-resolution regions were sampled at the mass resolution of the original MDPL2 simulations, using the GINNUNGAGAP code\footnote{GINNUNGAGAP generates initial conditions for zoom-in simulations by identifying the Lagrangian region of a halo and sampling it at the desired mass resolution while embedding it in the larger-scale volume. More information is available at: \url{https://ginnungagapgroup.github.io/ginnungagap/}}, while the mass of the gas particles is ($\rm M_{gas} $$\,= \, 2.36 \times 10^{8}\, h^{-1} \rm M_{\odot}$). The surrounding regions are represented with progressively lower-resolution particles, allowing the large-scale tidal field to be retained at lower computational cost.

    Among many other interesting contributions, \textsc{The300} has made significant efforts in obtaining accurate mass estimates and quantifying hydrostatic-equilibrium biases (\citealt{ansarifard_three_2020}; \citealt{li_et_al_2021}, \citeyear{li_et_al_2022}; \citealt{gianfagna_hydrostatic_2022}). The collaboration also investigates various aspects of galaxy clusters, including their relationships with connecting filaments (\citealt{kuchner_cosmic_2021}; \citealt{rost_et_al_2021}), cluster feeding (\citealt{kotecha_cosmic_2022}; \citealt{kuchner_cosmic_2021}), backsplash galaxies (\citealt{haggar_thethreehundred_2020}; \citealt{knebe_three_2020}), and the virial shock radius (\citealt{anbajagane_shocks_2022}; \citealt{baxter_shocks_2021}). Other hydrodynamic simulations further allow detailed studies of cluster properties such as density profiles (\citealt{mostoghiu_three_2019}; \citealt{li_three_2020}), substructure, baryon content (\citealt{arthur_span_2019}; \citealt{haggar_et_al_2021}; \citealt{Mostoghiu_et_al2021a}), dynamics, morphologies (\citealt{capalbo_three_2021}; \citealt{deluca_three_2021}; \citealt{zhang_span_2022}; \citealt{capalbo_et_al_2025}) and ICM thermalization (\citealt{sayers_et_al_2021}; \citealt{sereno_et_al_2021}). Additional research addresses the effects of mergers on the properties of the Brightest Cluster Galaxies (or BCGs; \citealt{contreras-santos_three_2022}) and the effect of environment on galaxy properties (\citealt{wang_three_2018}). The project further explores self-interacting dark matter and chameleon gravity, with continuous updates ensuring the robustness and breadth of the simulations.
    
    \begin{figure*}
     \centering
        \begin{subfigure}[b]{0.45\textwidth}
            \centering
            \includegraphics[width=7cm, height=7cm]{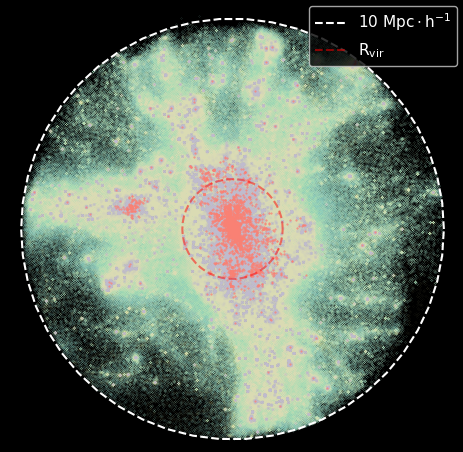}
        \end{subfigure}
        \begin{subfigure}[b]{0.45\textwidth}
            \centering
            \includegraphics[width=7cm, height=7cm]{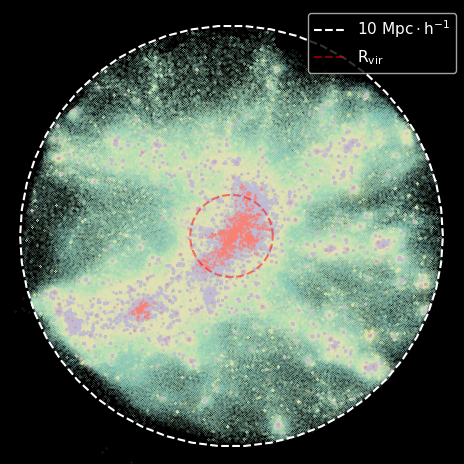}
        \end{subfigure}
    \caption{2D projection of the \texttt{NewMDCLUSTER\_0002} and \texttt{NewMDCLUSTER\_0006} clusters onto the XY plane at snapshot 117 (\( z = 0.276 \)). The figure shows the spatial distribution of dark matter and baryonic particles within each cluster. The stellar component is shown in pink, the gas component in purple, and the dark matter particles in green. The simulations were performed using the GIZMO-Simba code, a hybrid hydrodynamic and N-body solver designed for cosmological simulations. The clusters are centered in the simulation box for reference. The red dashed line indicates the virial radius (\( R_{\mathrm{vir}} \)) of each cluster.}
    \label{gizmo_sim}

    \end{figure*}

    We base our analysis on the run using the hydrodynamical code GIZMO-\textsc{Simba}. Our choice is mainly motivated by two considerations. First, the presence of baryons affects the mass profiles and the mass–concentration relation of halos, which are directly relevant for our analysis. Second, the GIZMO-\textsc{Simba} run provides galaxy colors and luminosities that are in better agreement with observations for the calibration adopted in \textsc{The300} \citep{cui_span_2022} Since an important part of our analysis focuses on evaluating systematic effects arising from magnitude limits and colour-based selection biases, this run is particularly well suited for the purposes of this work. We emphasize, however, that this choice is tailored to the specific goals of this study. Other runs within \textsc{The300} (either hydrodynamical or semi-analytic) may be more appropriate for different scientific objectives. The run was performed with the GIZMO code \citep{Hopkins2015} and the sub-grid galaxy formation models from the {\sc{Simba}} simulation \citep{dave_et_al_2019}, that  include radiative cooling, star formation and its associated feedback mechanisms, as well as the formation and growth of supermassive black holes. It also incorporates multiple modes of black-hole feedback, which are essential for accurately representing the impact of these massive objects on their host galaxies and the surrounding environment. The parameters of the {\sc{Simba}} models have been  recalibrated for the  \textsc{The300} runs as described in \cite{cui_span_2022} due to the lower numerical resolution of the runs compared to the original {\sc{Simba}} simulation. Figure ~\ref{gizmo_sim} shows a 2D projection of two clusters, highlighting the distribution of dark matter and baryonic components at $z = 0.276$.

\section{The caustic technique} \label{Caustic Technique}

    The caustic technique, as described by \cite{diaferio_infall_1997}, \cite{diaferio_mass_1999}, \cite{serra_measuring_2010}, and \cite{serra_identification_2013}, infers the mass profile of a galaxy cluster from the “trumpet-shaped” envelope (the caustics) traced by member galaxies in projected phase space ($\it r$, $\rm v_\mathrm{esc,los}(\it r)$, where los denotes the line-of-sight direction). The caustic amplitude, $\mathcal{A}(\it r)$,  approximates the escape velocity at each radius, $\rm v^{2}_{esc,los}(\it r)$, which allows one to infer the enclosed mass under the hypothesis of spherical symmetry: 
    
    \begin{equation}
    \hspace{+3em} GM(<r) - GM(r_{0}) = \int_{r_{0}}^{r}\mathcal{A}^{2}(x) \, \mathcal{F}_{\beta}(x) \, dx
    \end{equation}
    
    The function $\mathcal{F_{\beta}}$ is related to  the gravitational potential and the anisotropy profile $\beta = 1 - \left\langle \rm v_{\theta}^{2} + \rm v_{\phi}^{2} \right\rangle / 2\left\langle \rm v_{r}^{2} \right\rangle$. \citep{diaferio_infall_1997, serra_measuring_2010}.

    \begin{equation}
    \hspace{+8em} \mathcal{F}_{\beta} = -2G\pi g(\beta(r))\frac{\rho(r)r^{2}}{\Phi(r)}
    \end{equation}

    \begin{equation}
    \hspace{+9em} g(\beta(r)) = \frac{3 - 2\beta(r)}{1 - \beta(r)},
    \label{ec:anis_fuction}
    \end{equation}

    The projected $\rm v_\mathrm{esc}(\it r)$ surface is traced by applying kernel density estimation (KDE) techniques to the distribution of dynamical tracers in the $\it r$–$\rm v$ phase space. These tracers are subject to measurement uncertainties in both position and velocity in observational data. In this study we consider  spectroscopic errors of $\sim 10$ km s$^{-1}$ (equivalent to 0.12 $h^{-1}$ Mpc in normalized coordinates) and astrometric uncertainties of 0.006 $h^{-1}$ Mpc which are the values expected in the CATARSIS survey.  

    Following \cite{gifford_systematic_2013}, we adopt a fixed multi-dimensional Gaussian kernel for the KDE implementation, with the bandwidths along $\it r$ and $\rm v$ directions determined independently from the data sampling (see \citealt{silverman_1986}).

    \begin{equation}
    \hspace{+8em} K(r,\rm v) = \left(\frac{4}{3N}\right)^{1/5} \sigma_{r,\rm v},
    \end{equation}

    \noindent where $N$ is the number of dynamical tracers in the total phase-space, and $\sigma_{\it r,\rm v}$ is the dispersion along the radial or velocity dimensions. This adaptive kernel ensures that regions with few tracers are smoothed more heavily, while densely sampled regions are smoothed less, maintaining a balanced representation of the phase-space density.
       
    The ratio between the sizes of  the smoothing windows  in position and velocity, $q$, balances the relative contribution of these parameters in the identification of the caustic \citep{diaferio_infall_1997,geller_mass_1999}, and needs to be adjusted to the observational errors. We use the expected errors in velocity and position for the CATARSIS survey  and adopt   $q=21$. 
    Previous studies \citep[see, e.g.][]{geller_mass_1999, rines_infall_2000, rines_mass_2002} have shown that the derived cluster mass profile is relatively insensitive to the exact choice of $q$, as long as it remains within a reasonable range. In Section~\ref{Parameter space}, we study the impact of these parameters in the determination of the caustic.

    A physically motivated constraint is applied to the escape velocity profile, $\rm v_{\rm esc}(\it r)$. According to \cite{diaferio_mass_1999}, any realistic galaxy cluster model must satisfy:

    \begin{equation}
    \hspace{+9em} \frac{d \ln \rm v_{\rm esc}}{d \ln r} \le \zeta,
    \end{equation}

    \noindent where $\zeta$ limits how steeply the escape velocity can change with radius. The rationale is that very steep gradients in $v_{\rm esc}(\it r)$ would be unphysical, corresponding to unrealistic mass distributions. If an iso-density contour violates this constraint along its surface, the corresponding $\rm v_{\rm esc}(\it r)$ value is replaced with a corrected one.

    In practice, following \cite{serra_measuring_2010}, we adopt a looser limit of $\zeta = 2$ rather than the more conservative $\zeta = 1/4$ suggested by \cite{diaferio_mass_1999}. This choice allows greater flexibility in identifying the caustics in realistic cluster data, accounting for observational noise and sampling effects, while still ensuring that the inferred escape velocity profiles remain physically plausible.
    
\begin{figure}[htbp]
        \centering
        \begin{minipage}[b]{0.48\textwidth}
            \centering
            \includegraphics[width=8cm, height=8cm]{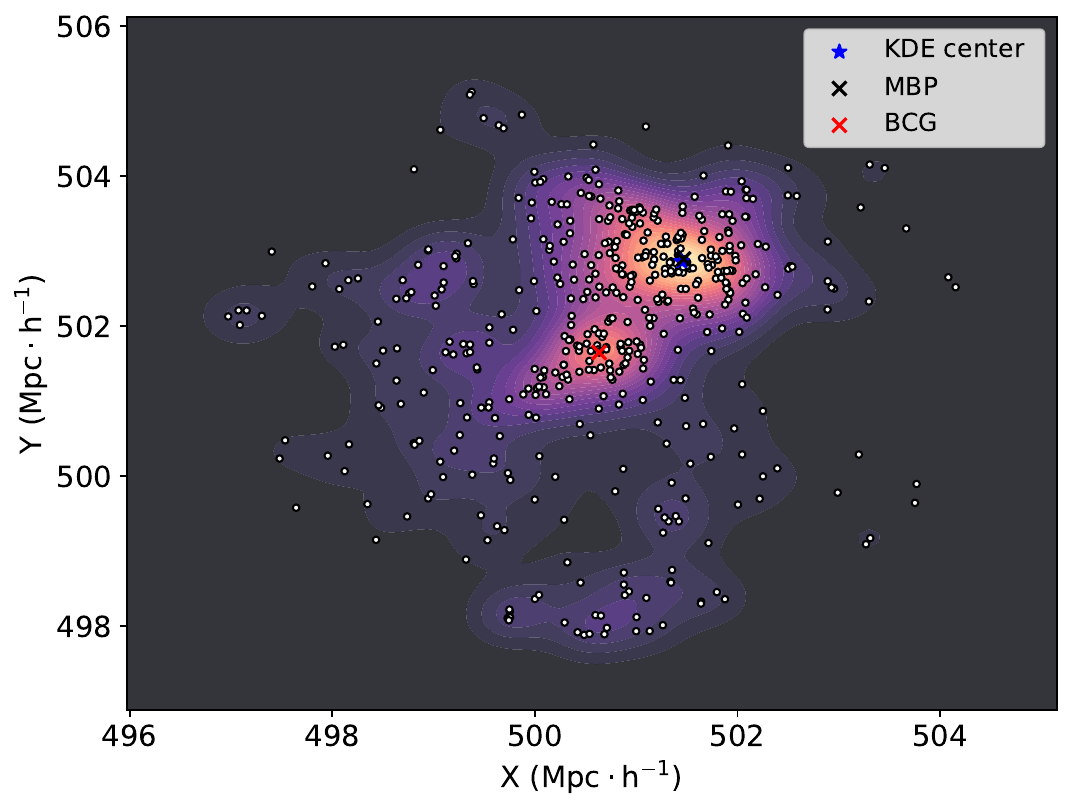}
        \end{minipage}%
        \hspace{0.04\textwidth} 
        \begin{minipage}[b]{0.48\textwidth}
            \centering
            \includegraphics[width=8cm, height=8cm]{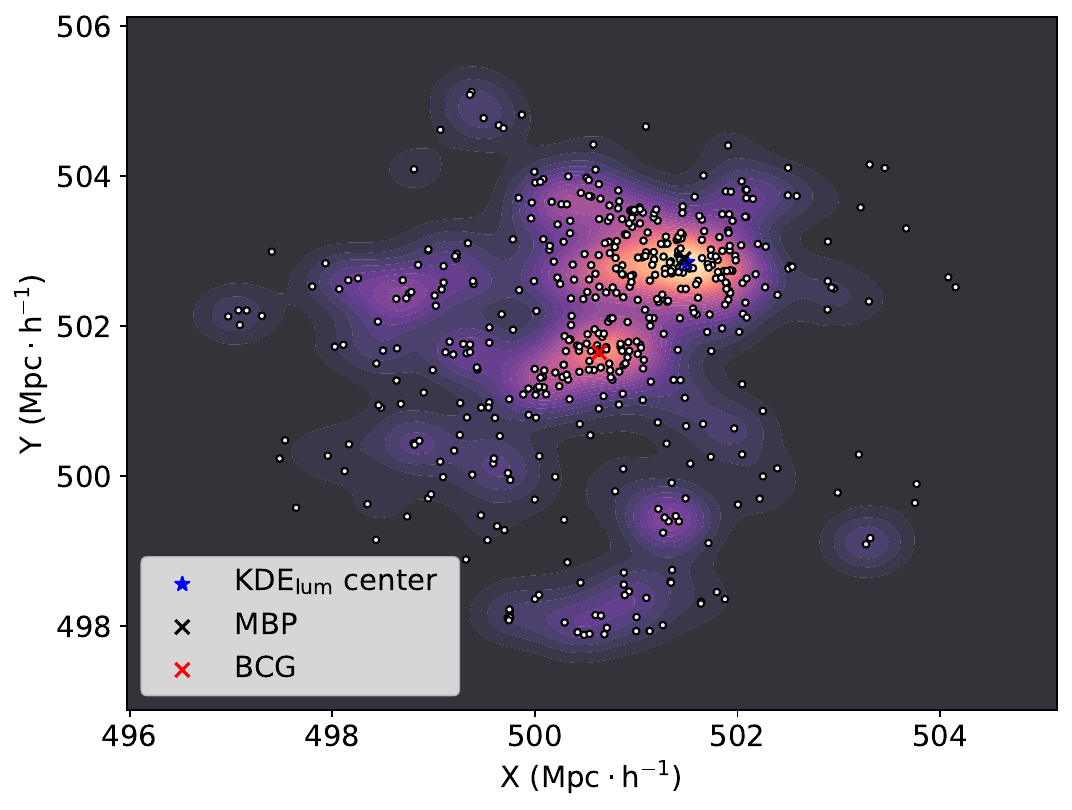}
        \end{minipage}
        \caption{Projected maps obtained with the KDE method for the cluster \texttt{NewMDCLUSTER\_0013} at redshift z=0.276. In the top panel only the positions of the galaxies have been used, while the bottom panel shows the results of weighting the positions with galaxy luminosities. The contours represent isodensity levels, with higher density regions indicated by the innermost lines. The star marks the center obtained from the KDE method, while the black cross indicates the BMP (Bound More Particle) center, identifying the most gravitationally bound region of the cluster, and the red cross shows the BCG center. The white dots correspond to galaxies in the cluster.} 
        \label{fig KDE}
    \end{figure}

    To derive cluster mass profiles with the caustic method, an accurate determination of the cluster center is required. In numerical simulations, the center of a halo is often defined as the minimum of the gravitational potential. However, alternative definitions are also used. For instance, some studies adopt the position of the most bound particle, which may be slightly offset from the potential minimum, while halo finders such as \textsc{AHF} \footnote{\url{http://popia.ft.uam.es/AHF}} \citep{KnollmannKnebe_2009} identify the center as the peak of an adaptively smoothed density field \citep[see, e.g.,][]{cui_halo_2016}. However, as we aim to assess biases relevant to forthcoming observations, we estimate the center using a KDE approach that  infers the projected surface density from the discrete positions  of observed galaxies. The cluster center is defined  as the location of the global maximum (mode) of the resulting density field.  When using only the coordinates, it is assumed that the number of halos, and not their mass, can trace the density field of the cluster. To check the influence of this assumption in the determination of the center, we have applied the method weighting with the luminosity of galaxies, as a proxy of mass. Figure \ref{fig KDE} shows an example of the position of the center determined with  the different methods. 
    
    The figure also displays the projected density map of a cluster obtained with both KDE implementations. In the luminosity-weighted case, brighter galaxies have a stronger influence on the density estimate, leading to a more concentrated distribution. However, the  inferred cluster centers in both cases differ by less than 2\% for the whole   sample.  The centers obtained with this method deviate only slightly from the position of the most bound particle. Across the full cluster sample, the median offset is 7\% for the standard KDE and 5\% for the luminosity-weighted KDE, with corresponding 84$^{\rm th}$-percentile values of 17\% and 14\%, respectively.
    This error in the center position has a negligible effect on the determination of the caustics \citep[see][]{serra_measuring_2010}. 

    We note that the results may depend on the choice of the line of sight due to the non-spherical nature of galaxy clusters. In this work, we adopt a single line of sight (along the simulation box Z-axis) to provide an assessment of the performance of the caustic method in The300 simulations from a representative set of observational data. A more detailed analysis, including projection effects along different sight-lines, substructure, rotation, is left for future work, where we aim to improve the caustic method beyond spherical assumptions. 
    
    The projected positions are used to build the surface density maps, while the line-of-sight velocity component, including the Hubble flow, is adopted to construct the redshift-space phase-space diagram. This setup provides a consistent and observationally motivated representation of each simulated cluster, serving as the basis for the application of the caustic technique. In this work, we use the galaxy catalogs provided by \textsc{Caesar}\footnote{\url{https://caesar.readthedocs.io/en/latest/}}, which include all galaxies identified within each cluster halo, without applying an additional membership selection. We note that, unlike simulations, the identification of true members in observations is considerably more uncertain; the impact of observational selection effects and field-of-view limitations is discussed in Sections~\ref{context of catarsis} and ~\ref{other_surveys}.

    \subsection{Impact of smoothing kernel size} \label{Parameter space}

    Although some studies have reported that the choice of smoothing kernel has little impact on the determination of density profiles (see, e.g., \citealt{rines_mass_2002}), we find that this effect varies from cluster to cluster and that, in a significant fraction of our sample, it exerts a non-negligible influence on the caustic determination. 
    The smoothing is typically parameterized by the scale of the position kernel $s$ and by the ratio between the velocity and position kernels $q$. 
    Although the typical positional uncertainties are $\sim 0.02~h^{-1}$ Mpc, $s$ is converted to km~s$^{-1}$ using the Hubble constant so that it can be directly compared with the velocity kernel when computing $q = h_v / h_r$. 
    Here, we investigate whether an optimal combination of these parameters exists that minimizes the errors in mass determination, and whether suitable values can be inferred for individual clusters based on their observed properties.

\begin{figure*}[ht!]
    \centering
    \begin{minipage}{0.68\textwidth}
        \centering
        \includegraphics[width=\linewidth]
        {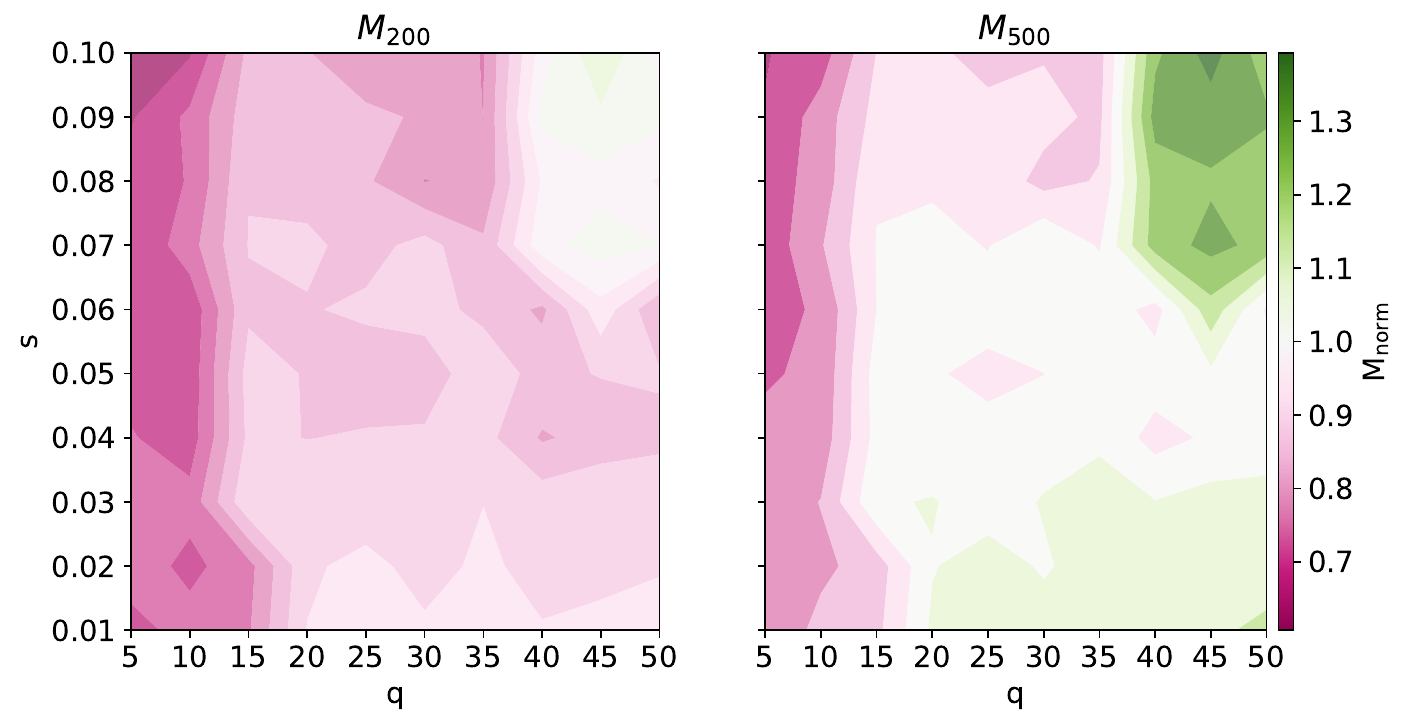}
    \end{minipage}
    \hfill
    \begin{minipage}{0.28\textwidth}
        \caption{
        Bi-parametric map of $s$ and $q$ for the cluster
        \texttt{NewMDCLUSTER\_0007} at $z = 0.276$.
        The color bars represent the ratio between the
        estimated and true values
        $M_{r}/M_{r,\mathrm{true}}$ for masses between
        $R_{200}$ ($M_{200}$) and $R_{500}$ ($M_{500}$).
        }
        \label{biparametric}
    \end{minipage}
    \end{figure*}

    We first explore the distribution of derived masses for different combinations of $s$ and $q$, with $s$ ranging from 0.01 to 0.1 in increments of 0.01 and $q$ from 5 to 50 in increments of 5.
    Figure~\ref{biparametric} shows, for one representative cluster, a two-parameter map of the masses within $r_{200}$ and $r_{500}$. As illustrated, several parameter combinations yield local minima, corresponding to better mass estimated, but these do not converge toward a well-defined region in parameter space. Moreover, the optimal combinations are not always consistent between the two mass estimated.

\begin{figure*}[ht!]
    \centering
    \begin{minipage}{0.68\textwidth}
        \centering
        \includegraphics[width=\linewidth]
        {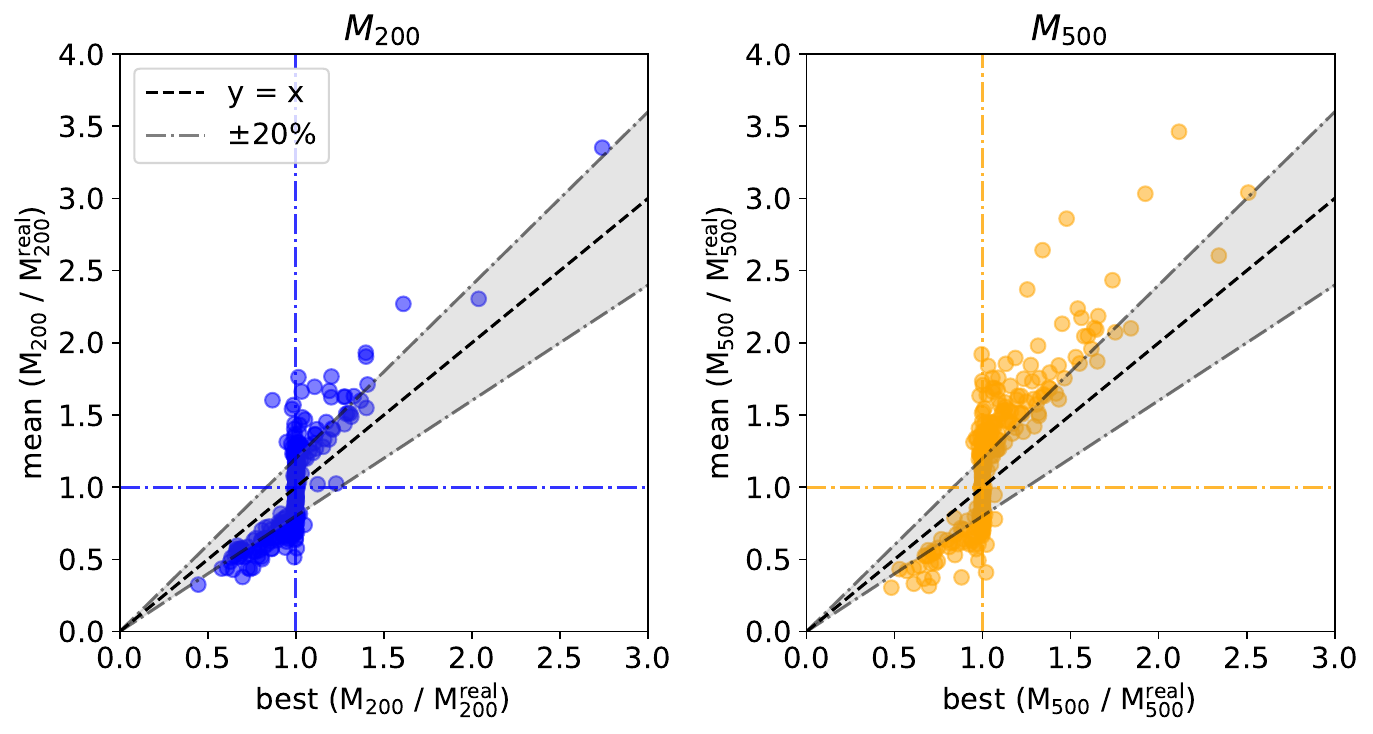}
    \end{minipage}
    \hfill
    \begin{minipage}{0.28\textwidth}
        \caption{
        Comparison between the best estimated values of
        $\sf M_{200}$ and $M_{500}$ (x-axis) and the average
        values computed using all combinations of $q$ and $s$
        explored in this work (y-axis). Each point corresponds
        to a cluster from the \textsc{The300} sample.
        The shaded region indicates deviations smaller than
        $\pm 20\%$ from the identity relation ($y=x$).
        }
        \label{best_parameters}
    \end{minipage}
    \end{figure*}

    To quantify the impact of these parameters, we compare the mean mass values obtained over the full grid of $(q, s)$ combinations with those derived from the optimal pair (i.e., the one yielding the smallest difference with respect to the true mass). Figure~\ref{best_parameters} illustrates these comparisons. 
    The deviation between the mean and optimal values provides an empirical measure of the robustness of the mass determination with respect to the kernel parameters: a small deviation indicates that the result is stable across the explored parameter space, while larger deviations reveal a stronger dependence on the particular parameter choice. 
    The 20\% threshold shown as the shaded region in Fig.~\ref{best_parameters} is adopted as a practical criterion to distinguish between robust and parameter-sensitive cases. This value roughly represents the typical variation observed across the cluster sample and serves as a convenient reference for identifying outliers whose mass estimates are particularly sensitive to the adopted parameters.

\begin{figure}[htbp]
        \centering
        \begin{minipage}[b]{0.48\textwidth}
            \centering
            \includegraphics[width=8.5cm, height=7cm]{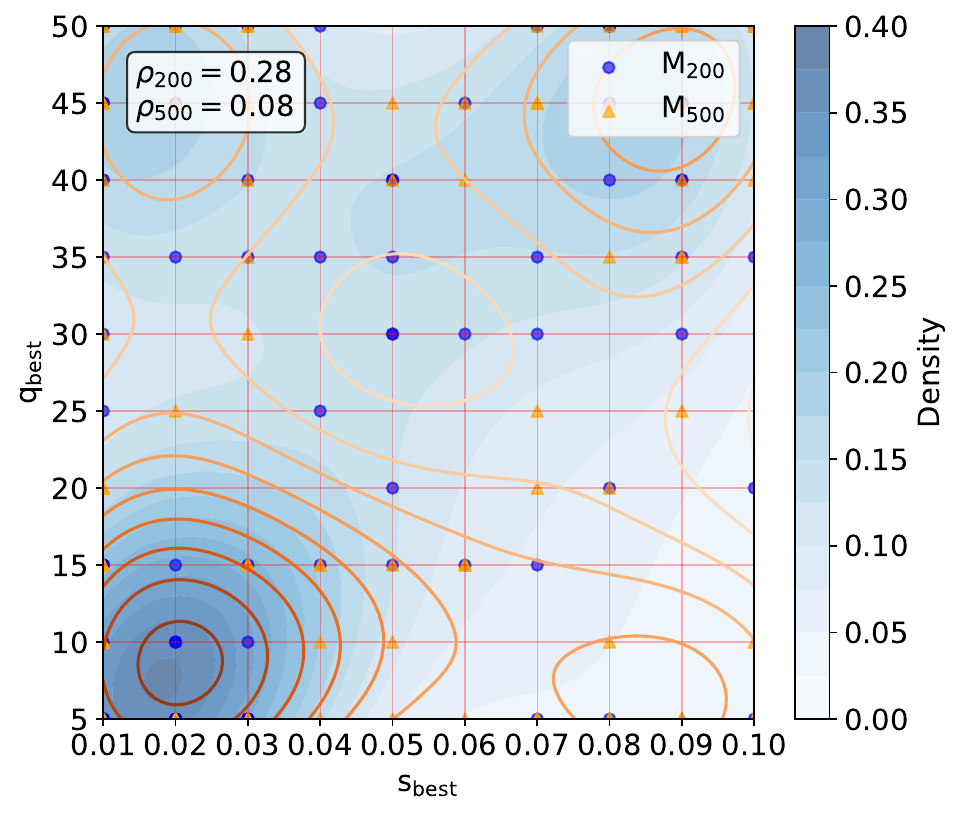}
        \end{minipage}%
        \caption{Distribution of the best-fit parameters in the $q$--$s$ plane for the cluster sample. Blue circles and orange triangles represent the values obtained using the $M_{200}$ and $M_{500}$ criteria, respectively. The shaded background and overlaid contours show the two-dimensional KDE of the parameter distribution, with the color scale indicating the local density across the parameter space. The Pearson correlation coefficients between $q$ and $s$ for each mass definition are showed in the inset.
        }
        \label{best_qs_m200_m500}
    \end{figure}

    Figure~\ref{best_qs_m200_m500} summarizes this analysis for all clusters in the sample. 
    For each cluster, we identified the combination of $(q, s)$ that minimizes the difference between the recovered and true mass, and the corresponding best-fitting values.
    Although regions of parameter space around this minimum can yield similarly good fits, we report the single best-fitting pair for clarity and consistency across the sample. 
    In general, smaller angular-distance kernel sizes yield better results when combined with a redshift kernel that is either five or fifty times larger, depending on the cluster. 
    However, the behavior varies significantly from cluster to cluster, and no clear trend can be established.

\begin{figure}[htbp]
    \centering
    \begin{subfigure}[b]{0.48\textwidth}
        \centering
        \includegraphics[width=8.5cm, height=7cm]{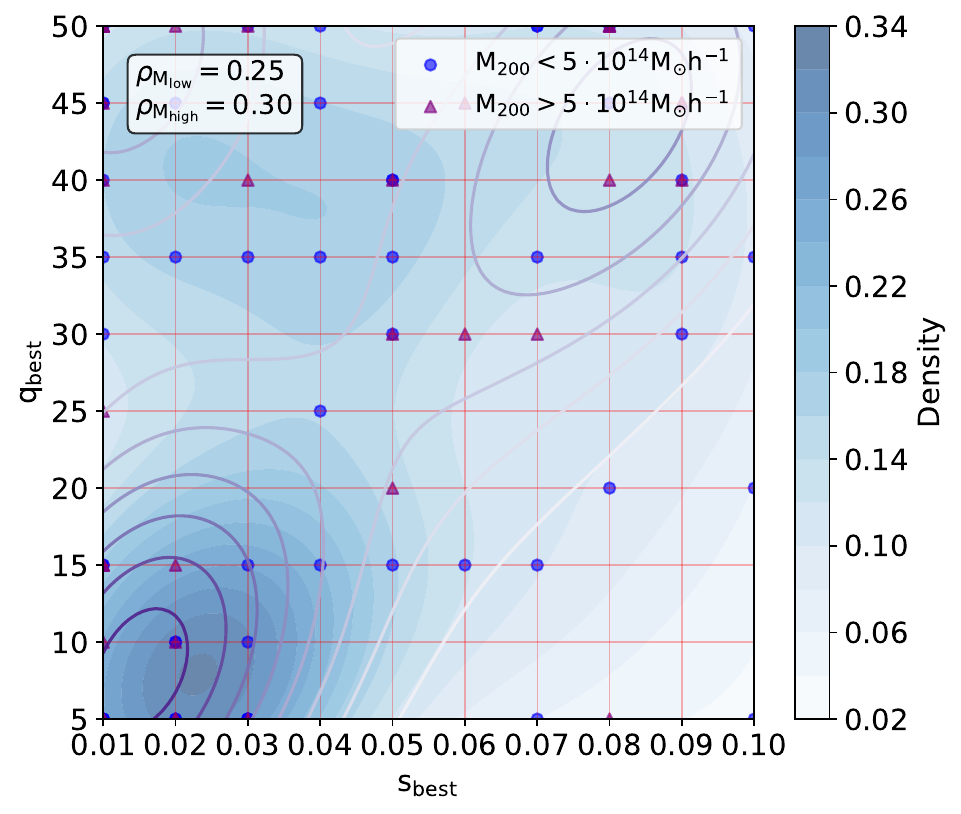}
        \caption{More vs. less massive clusters}
        \label{fig:best_qs_limM}
    \end{subfigure}
    \hfill
    \begin{subfigure}[b]{0.48\textwidth}
        \centering
        \includegraphics[width=8.5cm, height=7cm]{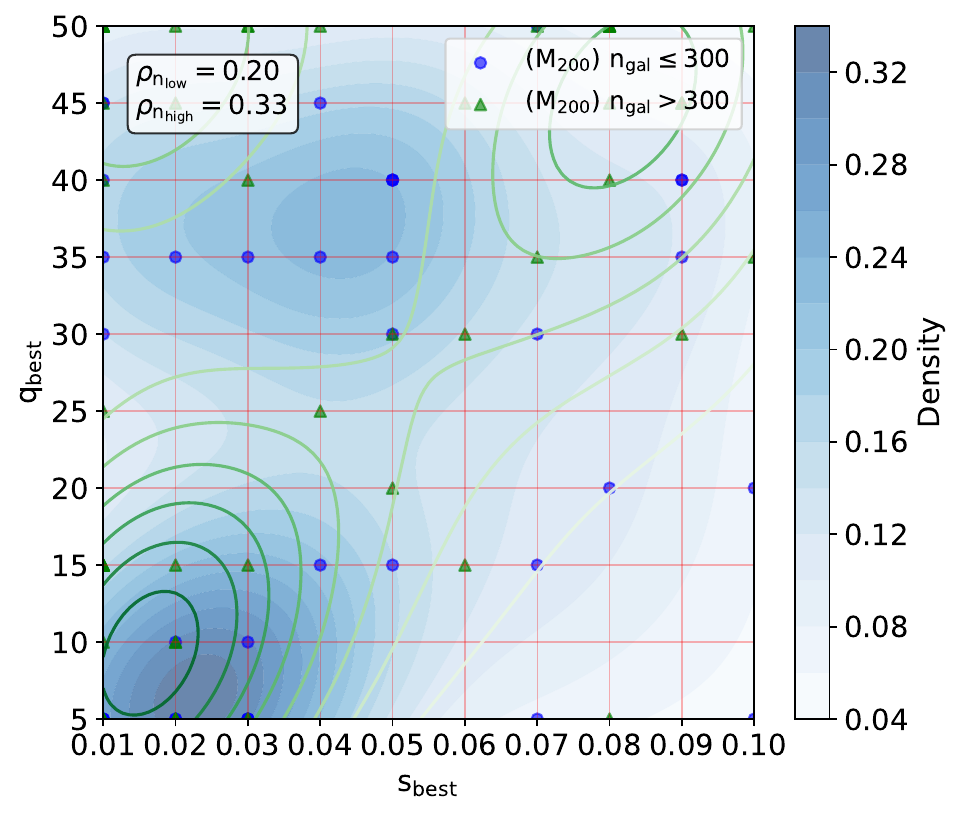}
        \caption{Same as Figure \ref{best_qs_m200_m500} but dividing the sample into clusters with more (green) and less (blue) than 300 members.}
        \label{fig:best_qs_limN}
    \end{subfigure}
    \caption{Same  as Figure~\ref{best_qs_m200_m500}, but dividing  the clusters into clusters with $\sf M_{200}$ larger ( triangles) and lower (circles) than $10^{14}$M$_{\odot}$h$^{-1}$, and in the lower panel according to the number of associated galaxies. This division allows us to investigate how the distributions of the best–fit values vary as a function of the total mass and the cluster richness.}
    \label{fig:qs_density_maps}
\end{figure}

    Finally, we examine possible relationships between the best $(q, s)$ combinations and other cluster properties, such as mass and richness. Figures~\ref{fig:best_qs_limM} and \ref{fig:best_qs_limN} show that the optimal smoothing kernels do not correlate with these parameters.Although the KDE distributions show a global minimum at lower values of q and s, several local minima are also present, corresponding to different parameter combinations. The optimal values therefore vary across clusters. We further quantify this by computing the correlation coefficients between the best-fit parameters (q, s) and cluster properties (mass and richness), finding no significant correlations ($\rho \leq 0.3$). Consequently, the current analysis does not allow us to identify a unique pair of $(q, s)$ values that performs optimally for all galaxy clusters.

\subsection{Impact of the anisotropy profile: an iterative method to improve the caustic determination} \label{Iterative Method}

    The velocity anisotropy profile, $\beta(r)$, plays a crucial role in the determination of the caustic amplitude (see Eq.~\ref{ec:anis_fuction}), as it governs the relation between the observed line-of-sight velocity distribution and the underlying escape velocity profile. Previous studies have often adopted simplified prescriptions, either assuming a constant anisotropy (e.g., \citealt{diaferio_mass_1999, gifford_systematic_2013}) or imposing a fixed radial dependence motivated by theoretical models or by averaging $\beta(r)$ over halo samples from numerical simulations (e.g., \citealt{mamon_dark_2005}). 
    
    All these  approaches, however, present important limitations. A constant anisotropy neglects the expected radial variation of galaxy orbits, while adopting a fixed functional form may introduce systematic biases if it does not accurately represent the dynamical structure of the specific cluster under study.

    To address these limitations, we adopt an iterative approach to estimate cluster masses. We begin by assuming a constant anisotropy and applying the caustic method to derive the mass and density profiles. The concentration parameter is then obtained by fitting a Navarro–Frenk–White (NFW) profile to the inferred density distribution. 
    
\begin{figure}[htbp]
        \centering
        \begin{minipage}[b]{0.48\textwidth}
            \centering
            \includegraphics[width=8cm, height=6cm]{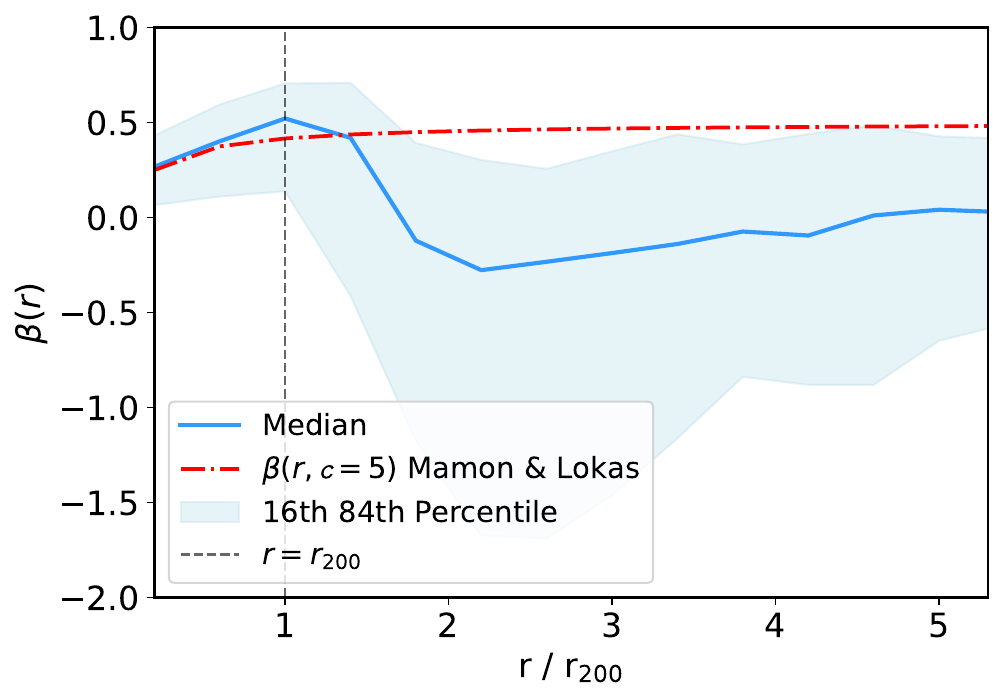}
        \end{minipage}%
        \caption{Anisotropy profile derived from the theoretical definition of the anisotropy parameter $\beta(r)$ for the whole sample (blue line). The thick line shows the median profile across all clusters, while the shaded areas represent the 16$^{\rm th}$ and 84$^{\rm th}$ percentiles, respectively. The red line corresponds to the anisotropy profile from \citep{mamon_dark_2005} with a concentration parameter fixed to $c = 5$. In future work, we will explore the differences arising when clusters exhibit rotational effects, which can cause the two definitions of $\beta(r)$ to diverge. In this work, we compute $\beta(r)$ using the radial and tangential velocity dispersions ($\sigma_r$ and $\sigma_t$) in each radial bin, instead of the ideal definition based on the mean squared velocities. This approach is appropriate for clusters without significant rotational support, where both definitions coincide.}
        \label{fig anisotropy profile}
    \end{figure}

    Using these initial results, we recompute the caustic profile adopting two radially varying anisotropy models: (i) a data-driven profile derived directly from the definition of $\beta(r)$, and (ii) the analytical prescription proposed by \citet{mamon_dark_2005}. The empirical $\beta(r)$ profile is constructed by dividing the galaxies into logarithmically spaced radial bins from the cluster center, considering all bound galaxies within $R_{\rm max}$, and computing the ratio between the tangential and radial velocity dispersions in each bin (see Fig.~\ref{fig anisotropy profile}).  

\begin{figure}[htbp]
        \centering
        \begin{minipage}[b]{0.48\textwidth}
            \centering
            \includegraphics[width=8cm, height=6cm]{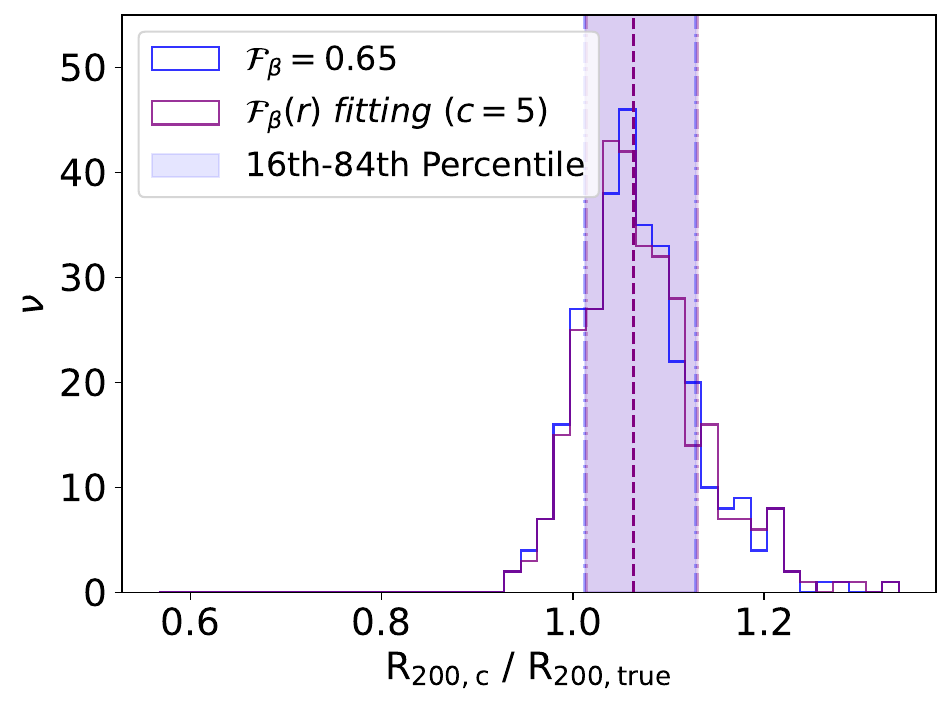}
        \end{minipage}%
        \hspace{0.04\textwidth} 
        \begin{minipage}[b]{0.48\textwidth}
            \centering
            \includegraphics[width=8cm, height=6cm]{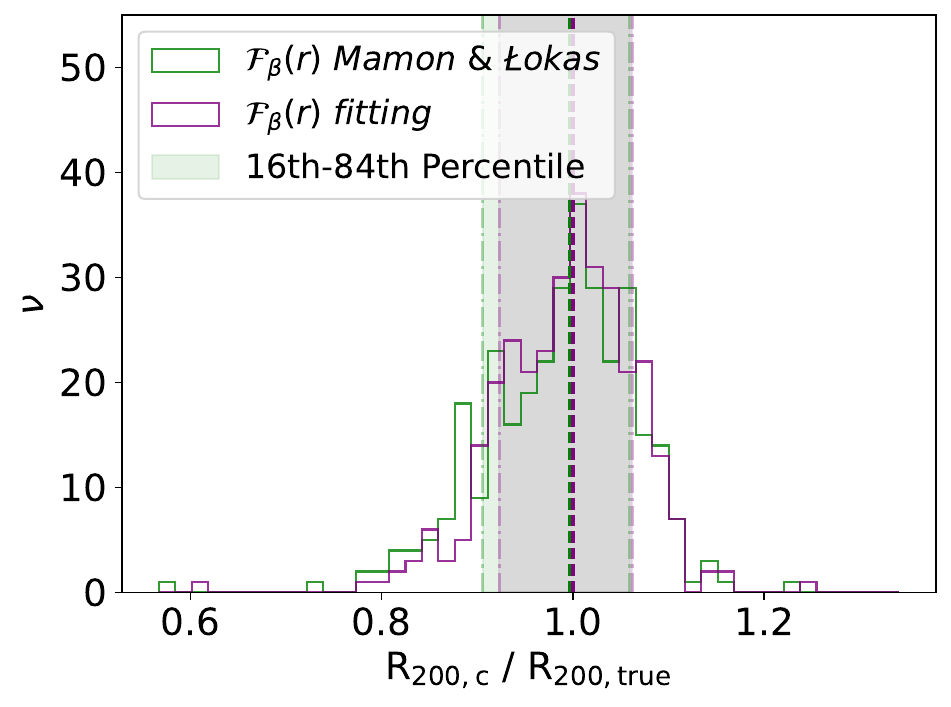}
        \end{minipage}
        \caption{
        All panels show the ratio between the derived and true \(R_{200}\) values. The top panels correspond to the non-iterative approach, comparing a constant \(F_{\beta}\) profile (blue) and a Mamon \& Łokas anisotropy model with fixed concentration \(c=5\) (purple). The bottom panels show the results obtained using the iterative method, comparing a Mamon \& Łokas profile (green) and a fitted \(F_{\beta}\) model (purple). In all panels, bold lines indicate the median values, while shaded regions correspond to the 16$^{\mathrm{th}}$ and 84$^{\mathrm{th}}$ percentiles. For the non-iterative approach, we obtain a median ratio of \(1.06^{+0.07}_{-0.05}\), indicating a systematic overestimation. In contrast, the iterative method yields a median ratio of \(1.00^{+0.06}_{-0.09}\). 
    }
        \label{fig hist_radius}
    \end{figure}

\begin{figure}[htbp]
        \centering
        \begin{minipage}[b]{0.48\textwidth}
            \centering
            \includegraphics[width=8cm, height=6cm]{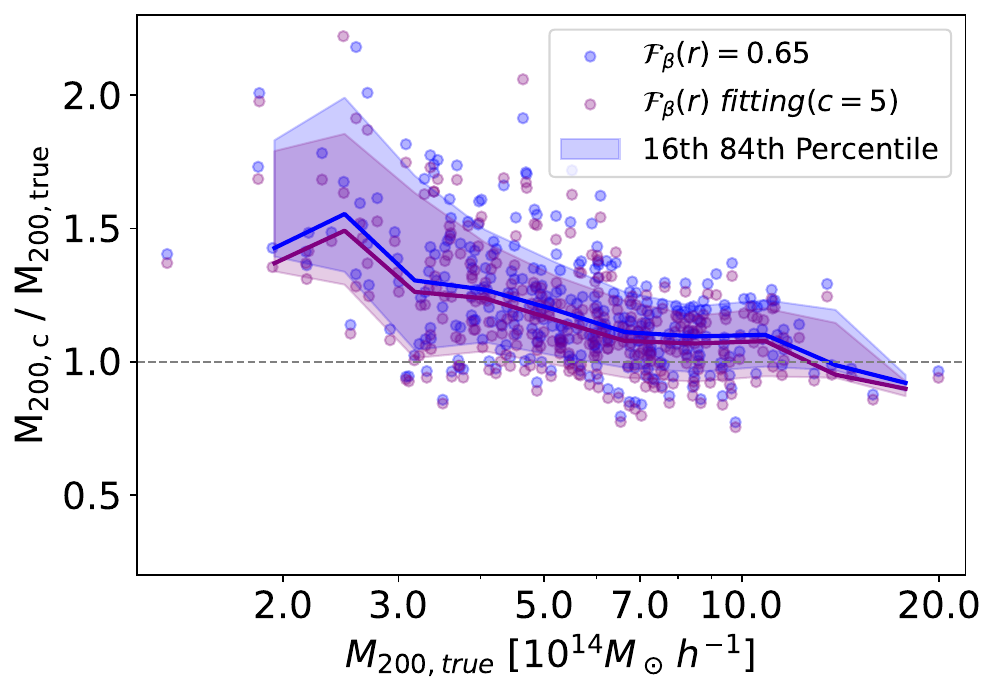}
        \end{minipage}%
        \hspace{0.04\textwidth} 
        \begin{minipage}[b]{0.48\textwidth}
            \centering
            \includegraphics[width=8cm, height=6cm]{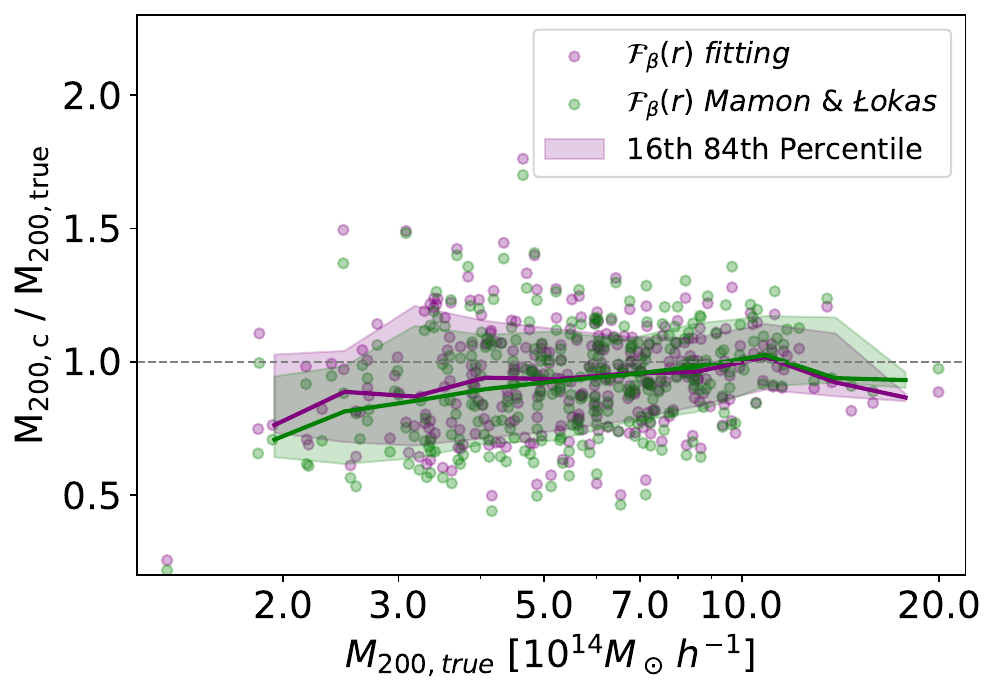}
        \end{minipage}
        \caption{
        Comparison of the ratio between the caustic mass estimate and the profile mass, $\rm M_{200,c} / \rm M_{200, true}$, as a function of $\rm M_{200}$, for two different approaches. Top panel: $\mathcal{F}_{\beta}(r)$ fitting with fixed parameters $\mathcal{F}_{\beta} = 0.65$ (blue) and concentration $c = 5$ (purple). Bottom panel: $\mathcal{F}_{\beta}(r)$ fitting with free parameters (purple) compared to the Mamon \& Łokas method (green). In both panels, points show individual measurements, solid lines represent the median ratio in logarithmic mass bins, and shaded regions indicate the 16$^{\rm th}$ to 84$^{\rm th}$ percentile range. The gray dashed horizontal line marks the unity ratio, $\rm M_{200, c} = \rm M_{200,true}$.}
        \label{fig mass}
    \end{figure}

    Both anisotropy profiles provide a good description up to $R_{\mathrm{200}}$. 
    Beyond this radius, the theoretical profile shows a decrease that coincides with the transition between the virialized region and the surrounding infall zone, where galaxy orbits become increasingly radial. 
    In this outer region, the number of tracer galaxies per radial bin is smaller, which increases the statistical uncertainty of $\beta$ but does not systematically bias its value. 
    At larger radii, the profile rises again as the number of sources increases and the infall motion becomes more dominant. 
    Figures~\ref{fig hist_radius} and~\ref{fig mass} show the resulting $R_{\mathrm{200,c}}$ and $M_{200,c}$ obtained with the iterative caustic method.

    As can be seen, the results obtained when applying the iterative method better reproduce the true values. For the full sample of 324 clusters, the non-iterative approach systematically overestimates the recovered masses. Using a constant \(F_{\beta}=0.65\), we obtain a median ratio \(M_{200,\mathrm{c}}/M_{200,\mathrm{true}}=1.16\), corresponding to a bias of \(+16\%\) with a scatter of \(0.23\). When adopting the fitted \(F_{\beta}\) profile (\(c=5\)), the median ratio decreases to \(1.12\), with a bias of \(+12\%\) and a scatter of \(0.22\). In contrast, the iterative method significantly reduces both the bias and the scatter. Using the fitted \(F_{\beta}\) profile, we obtain \(M_{200,\mathrm{c}}/M_{200,\mathrm{true}}=0.94\), with a bias of \(-6\%\) and a scatter of \(0.18\). Using the Mamon \& Łokas anisotropy model, the recovered ratio is \(0.93\), with a bias of \(-7\%\) and a scatter of \(0.2\). This improvement is particularly relevant for the least massive systems, where the non-iterative method can overestimate the masses by factors of \(\sim 1.5\)--\(2\). For the following results, we therefore adopt the iterative method together with our anisotropy profile.

\section{Bias in the mass determination due to sample selection} \label{context of catarsis}

     In observational studies of galaxy clusters, the selection of galaxies used to trace the gravitational potential can introduce biases in the mass determination. These biases arise from various factors, including the preferential use of red-sequence galaxies, magnitude-limited samples, and line-flux cuts, which can systematically over- or under-estimate cluster masses. In this section, we investigate how different selection criteria affect the mass estimates obtained with the caustic method. Specifically, we examine: (i) the impact of restricting the analysis to Red-Sequence galaxies, and (ii) the effects of observational limits in apparent magnitude and H$\alpha$ flux, which influence the detectability of galaxies with different star formation rates. Our goal is to quantify these biases and to assess how the planned CATARSIS survey, with its deep, unbiased spectroscopic observations, will mitigate them.

\subsection{Red sequence}

\begin{figure}[htbp]
        \centering
        \begin{minipage}[b]{0.48\textwidth}
            \centering
            \includegraphics[width=8cm, height=6cm]{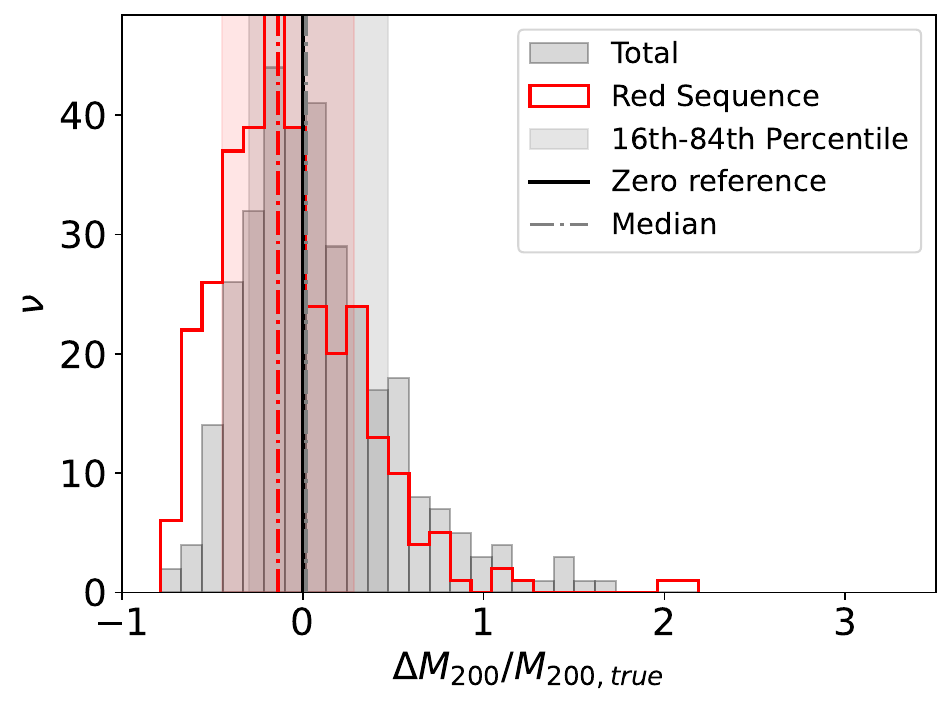}
        \end{minipage}%
        \caption{Relative error in the determination of $M_{200}$ using the caustic method, for the full galaxy sample (grey) and for the RS sample (red). Shaded areas indicate the 16$^{\mathrm{th}}$ and 84$^{\mathrm{th}}$ percentiles of the distributions; dashed lines show the median values, and the black line marks the reference value of zero. For the full sample we obtain $0.02^{+0.45}_{-0.32}$, while for the RS sample we find $-0.14^{+0.42}_{-0.30}$.}
        \label{fig RS}
    \end{figure}

    In spectroscopic surveys of galaxy clusters, it is common practice to prioritize galaxies lying on the red sequence (RS). Selecting this  population, clearly identifiable as a tight locus in color–magnitude space, helps to reduce contamination from foreground and background objects, improves the efficiency of spectroscopic target selection.  However,  focusing solely in these galaxies, this method can introduce a bias that typically leads to an underestimation of the true cluster mass as these objects are abundant near the dense cluster central regions\footnote{The higher relative abundance of luminous Red-Sequence galaxies in the center of the clusters is a direct consequence of the well-known morphology-density relation first found by \cite{dressler_1980}.}. To quantify this effect, we isolate Red-Sequence galaxies as those located within a $\pm 5\%$ interval of our best-fitting Red Sequence in the $m_{g}-m_{r}$ vs$.$ $m_{r}$ color-magnitude diagram. Besides objects in the Blue Cloud and the Green Valley, this strategy also eliminates galaxies that appear above the Red Sequence because of dust-reddened star formation. Once the Red Sequence is defined, we calculate the cluster mass using the caustic method. As shown in Fig.~\ref{fig RS}, the masses derived from this approach are consistently lower than the actual values, and we achieve more accurate results when the method is applied without pre-selecting Red-Sequence galaxies.

\subsection{$ H\alpha$ flux and magnitude limits} \label{Limits in magnitude and flux}

    This subsection investigates the impact of sample selection, specifically magnitude and H$\alpha$ flux limits, on the determination of cluster masses using the caustic method. The  H$\alpha$ fluxes  are derived from the star formation rates (SFRs) using the standard conversion of \citet{kennicutt_1998}.

\begin{figure}[htbp]
        \centering
        \begin{minipage}[b]{0.48\textwidth}
            \centering
            \includegraphics[width=8cm, height=6cm]{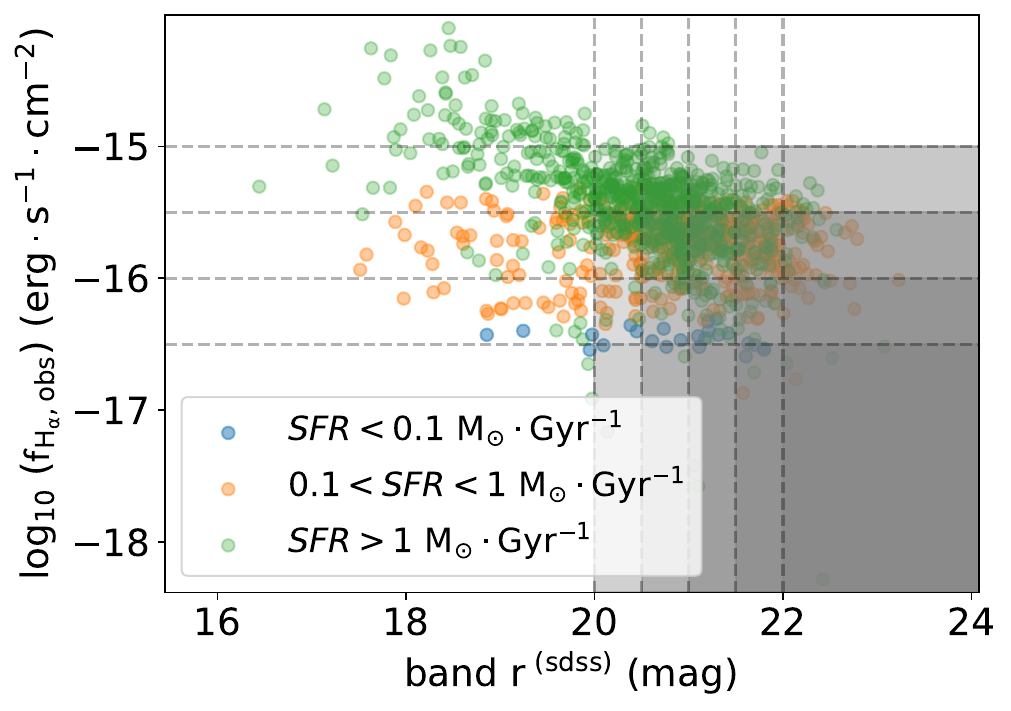}
        \end{minipage}%
        \caption{Magnitude and H$\alpha$ flux distributions for the cluster \texttt{NewMDCLUSTER\_0001} at snapshot 117 ($z = 0.276$). Galaxies are colored according to three star formation rate (SFR) bins: low (blue), medium (orange), and high (green), illustrating how star formation activity correlates with both $r$-band magnitude and H$\alpha$ flux. The shaded regions indicate the magnitude and H$\alpha$ flux cuts adopted for the analysis, with the magnitude selection performed in the $r$-band.}
        \label{fig flux_mag}
    \end{figure}

    Figure~\ref{fig flux_mag} illustrates, for a representative cluster, the relation between H$\alpha$ flux and (uncorrected) apparent magnitude in the SDSS $r$ band. The shaded regions indicate the adopted selection cuts in magnitude and H$\alpha$ flux, defined according to the specifications of the TARSIS integral field spectrograph and the current observational strategy of the CATARSIS survey \citep{Armando_2024}.

    Galaxies are further classified according to their SFR into three regimes: $\mathrm{SFR} > 0.1,\mathrm{M_\odot,yr^{-1}}$, $0.01 < \mathrm{SFR} \leq 0.1,\mathrm{M_\odot,yr^{-1}}$, and $\mathrm{SFR} \leq 0.01,\mathrm{M_\odot,yr^{-1}}$. As expected, galaxies with lower SFRs tend to exhibit weaker H$\alpha$ emission and fainter apparent magnitudes, making them more challenging to detect within the survey limits. Consequently, these systems are preferentially excluded by the adopted cuts. In contrast, galaxies with higher SFRs are more easily detected and remain well represented in the selected sample, although some incompleteness persists due to the imposed observational thresholds.

\begin{figure}[htbp]
        \centering
        \begin{minipage}[b]{0.48\textwidth}
            \centering
            \includegraphics[width=7cm, height=6cm]{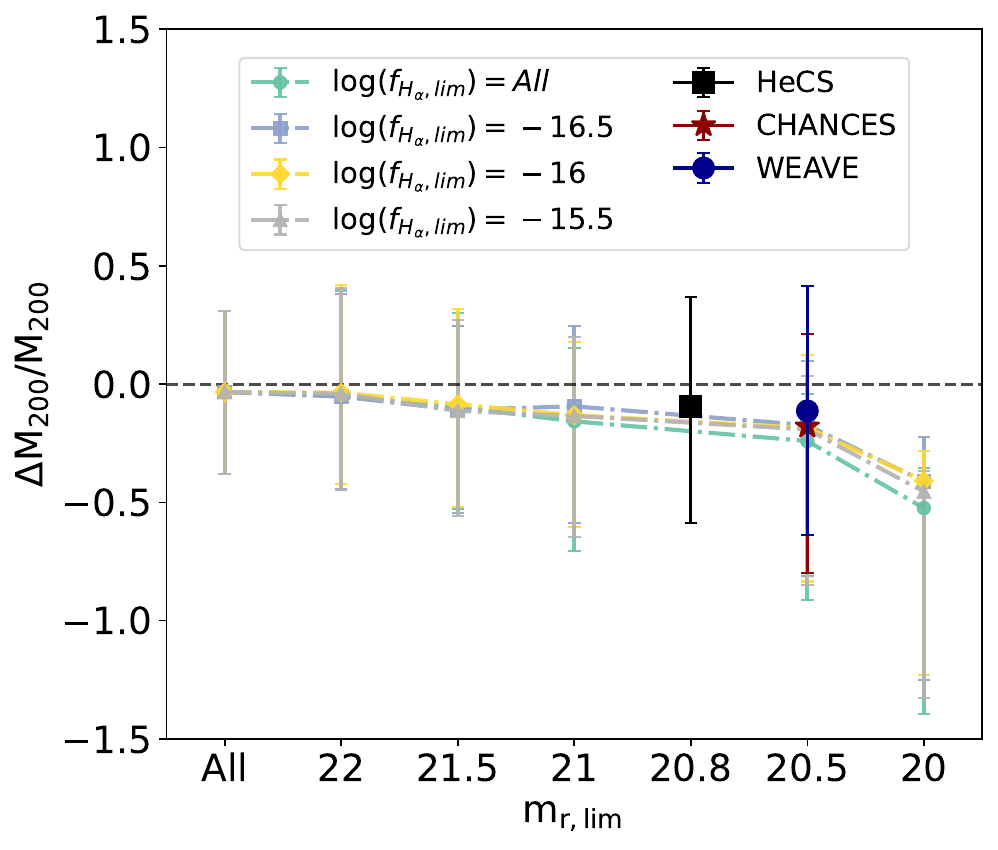}
        \end{minipage}%
        \caption{Relative $M_{200}$ values obtained for different cuts in $r$-band magnitude and H$\alpha$ flux. The y-axis shows the values normalized to the true cluster mass, while the x-axis represents the magnitude cuts. Colors indicate the H$\alpha$ flux cuts in logarithmic scale. A value of zero corresponds to perfect agreement between the calculated and true cluster mass. Error bars indicate the 16$^{\rm th}$ and 84$^{\rm th}$ percentiles of the distribution across \textsc{The300} clusters.}
        \label{fig flux_mag_comp}
    \end{figure}

    Regarding our observations cuts in line-flux and apparent (continuum) magnitude, we chose $\log(f_{H\alpha, \rm lim} [\rm erg\,s^{-1}\,cm^{-2}])$ between $-$16.5 and $-$15.5 and $m_{r,\rm lim}$ from 22 to 20.5\,mag. Figure~\ref{fig flux_mag_comp} shows the results of $\sf M_{200}$ that we have obtained for the whole set of \textsc{The300} clusters. As we cut in apparent magnitude, the average value shifts to higher values, resulting in an overestimation of the actual cluster masses. As for the cuts in flux, as we exclude galaxies with progressively fainter H$\alpha$ flux, we obtain greater dispersion compared to the real (true) mass values.
    
\section{Comparison with other surveys} \label{other_surveys}

In this section, we evaluate the expected biases in the determination of the caustic masses with other ongoing and upcoming spectroscopic surveys of galaxy clusters. The analysis highlights the differences in redshift range, spectroscopic depth, sampling strategy, and spatial coverage, which ultimately determine the feasibility of dynamical mass measurements based on galaxy redshifts.  

\subsection{WEAVE cosmological clusters survey}  
The WEAVE cluster survey \citep{jin_wide-field_2024} will soon target approximately 70 SZ-selected clusters at $z < 0.5$, using both its LIFU (for $z > 0.25$) and MOS (for $z < 0.25$) configurations with the low-resolution setup ($R \approx 5000$). The MOS mode covers a wide $2^\circ$ diameter field-of-view (FoV) ideal for low redshift (corresponding to $\sim 3.9$ kpc$\,/$arcsec at $z = 0.25$), while the LIFU mode provides a significantly smaller coverage of $90'' \times 78''$ ($\sim 6.2$ kpc$\,/$arcsec at $z = 0.5$). The limiting magnitude is not explicitly provided, but values of $r \approx 20.5$--21 at $z = 0.25$ are a reasonable assumption.  

The large FoV in physical units at $z < 0.25$ ensures adequate sampling for dynamical analyses, while at higher redshift the limited LIFU coverage restricts observations to the cluster core, preventing the application of the caustic method.  

\subsection{Hectospec Cluster Survey (HeCS) and HeCS-SZ}  
The Hectospec Cluster Survey \citep{rines_hecs-sz_2016} includes 58 X-ray selected clusters at $0.1 < z < 0.3$, while its SZ extension (HeCS-SZ) targets 21 systems at $0.05 < z < 0.20$. The survey employs a $1^\circ$ diameter FoV, corresponding to radial scales from $\sim 1$ Mpc at $z = 0.05$ to $\sim 4.5$ Mpc at $z = 0.3$. The magnitude limit is $r = 20.8$, with typical sampling of $\sim 300$ galaxies per cluster and velocity uncertainties around $50 \,\mathrm{km\,s^{-1}}$.  

The combination of depth, wide FoV, and high spectroscopic completeness of HeCS makes it one of the most suitable surveys for the caustic method — a fact demonstrated by \cite{rines_measuring_2013}, who obtained mass profiles out to large radii for 58 X‑ray selected clusters; HeCS masses have since been widely used as a reference for cluster mass calibrations and studies of mass accretion (e.g., \cite{pizzardo_mass_2021, maughan_hydrostatic_2016, Andreon_2016}.
 
\subsection{4MOST galaxy cluster surveys: CHANCES and eROSITA follow-up}

The CHANCES Evolution program \citep{sifon_chances_2025} targets $\sim 50$ massive clusters in the range $0.07 < z < 0.45$, selected above $M_{200} > 7 \times 10^{14}\,M_\odot$. Spectroscopic coverage extends out to $5 \times r_{200}$, ensuring radial completeness well into the cluster outskirts. The limiting magnitude is $r < 20.5$, providing sufficient sampling depth for member identification. The combination of large radial coverage, depth, and focus on massive systems makes CHANCES highly suitable for caustic-based analyses, with the potential to complement and extend results from HeCS to higher redshift.

The eROSITA Galaxy Cluster Survey \citep{4most_2019_e_ROSITA} will provide redshifts for systems up to $z < 0.7$, targeting galaxies with $K < 18$ (roughly $r \approx 19$). Current estimates indicate magnitude limits of $z_\mathrm{DECam} \leq 21.5$ for cluster members and $r_\mathrm{DECam} \leq 18.5$ for the filament survey (with no color cut). Given the low number of spectroscopic members per cluster, the full caustic methodology will only be applicable to a subset of clusters, requiring stacking of multiple systems to obtain reliable dynamical measurements.

\subsection{Comparison of cluster surveys for caustic mass reconstruction}

From this comparison, we conclude that HeCS/HeCS-SZ and CHANCES offer the most favorable conditions for applying the caustic method, due to their wide spatial coverage and high spectroscopic completeness. The results obtained from the CHANCES and HeCS surveys are consistent with the findings reported in Section \ref{Limits in magnitude and flux} (see Figure \ref{fig flux_mag_comp}). WEAVE provides sufficient sampling only for low-redshift systems ($z < 0.25$ in MOS mode), while its higher-redshift LIFU configuration is limited to the cluster core, making caustic mass reconstruction unfeasible. In contrast, the eROSITA Galaxy Cluster Redshift Survey, despite its cosmological reach, does not provide the necessary sampling per cluster and therefore does not constitute a practical dataset for caustic analyses. Finally, CATARSIS, being able to reach galaxies down to $m_r$$\simeq$21, will allow the study of fainter cluster members, offering a depth comparable to other surveys based on pre-imaging but adding galaxies selected by line emission down to $\log(f_{H\alpha, \rm lim} [\rm erg\,s^{-1}\,cm^{-2}])$=$-$16.5 , therefore establishing itself as a promising option for applying the caustic method with minimal mass bias and cluster-by-cluster dispersion.

\section{Conclusions} \label{conclusions}

In this work, we have assessed the reliability of the caustic technique for recovering galaxy cluster mass profiles, using \textsc{The300} simulations as a preparation for the forthcoming CATARSIS survey. Our study demonstrates that the standard method suffers from systematic biases, but these can be significantly reduced through a refined implementation.  

We introduced an iterative approach that combines anisotropy profile fitting with the caustic mass reconstruction, improving the agreement with true cluster masses. We find that the method exhibits a systematic trend with halo mass, with biases in the low-mass regime reaching values of order ~40\%, depending on the modelling assumptions.
This behaviour is consistent with previous studies of the caustic technique in both numerical and observational contexts, which have shown that mass estimates are generally unbiased on average but subject to significant scatter driven by projection effects, limited phase-space sampling, and uncertainties in the velocity anisotropy profile \citep{serra_measuring_2010, gifford_systematic_2013, armitage_cluster-eagle_2019, pizzardo_illustristng_2023}.
In particular, the iterative scheme reduces the systematic deviation, leading to an overall improvement in the mass calibration of order ~20\% when compared to the non-iterative case. However, we note that the inferred masses remain sensitive to the assumed velocity anisotropy profile, consistent with the well-known mass–anisotropy degeneracy in caustic reconstructions.

The analysis further highlights the critical role of galaxy sampling. Restricting measurements to Red-Sequence galaxies consistently underestimates cluster masses, while the blind integral-field spectroscopy of CATARSIS ensures a more complete representation of the cluster population and eliminates this bias. The influence of magnitude and H$\alpha$ flux limits was also investigated: bright magnitude cuts tend to overestimate masses, whereas relatively bright flux thresholds increase the scatter. The planned depth of CATARSIS strikes a balance between these effects, providing both precision and statistical robustness.  

When compared with other spectroscopic surveys, CATARSIS emerges as uniquely powerful thanks to its combination of depth and wide spatial coverage. It will deliver dynamical mass profiles of comparable quality to those from benchmark surveys such as HeCS, while also offering an essential complement to weak-lensing studies. Overall, CATARSIS will provide a decisive step forward in the precise calibration of galaxy cluster masses, thereby enhancing the cosmological constraints that can be derived from these structures.

\begin{acknowledgements}
This work acknowledges support from the following grants funded by MCIN/AEI/10.13039/501100011033/FEDER, EU: PID2022-138621NB-I00 (TARSIS) and PID2022-138855NB-C31 (CATARSIS). This work has also been made possible by the 'The Three Hundred' collaboration.\footnote{\url{https://www.the300-project.org}}. A. Knebe acknowledges support from the Ministerio de Ciencia e Innovación (MICINN) under research grant PID2021-122603NB-C21 as well as by project PID2024-156100NB-C21 financed by MICIU /AEI/10.13039/501100011033 / FEDER, UE. He further thanks The Cranberries for 'zombie'.
\end{acknowledgements}

\bibliography{biblio}

@article{dave_et_al_2019,
       author = {{Dav{\'e}}, Romeel and {Angl{\'e}s-Alc{\'a}zar}, Daniel and {Narayanan}, Desika and {Li}, Qi and {Rafieferantsoa}, Mika H. and {Appleby}, Sarah},
        title = {SIMBA: Cosmological simulations with black hole growth and feedback},
      journal = {MNRAS},
         year = 2019,
       volume = {486},
       number = {2},
        pages = {2827--2849},
          doi = {10.1093/mnras/stz937}
}

@article{dressler_1980,
       author = {{Dressler}, A.},
        title = {Galaxy morphology in rich clusters: implications for the formation and evolution of galaxies},
      journal = {ApJ},
         year = 1980,
       volume = {236},
        pages = {351--365},
          doi = {10.1086/157753}
}

@article{cui_span_2022,
       author = {Cui, Weiguang and Dave, Romeel and Knebe, Alexander and Rasia, Elena and Gray, Meghan and Pearce, Frazer and Power, Chris and Yepes, Gustavo and Anbajagane, Dhayaa and Ceverino, Daniel and Contreras-Santos, Ana and de Andres, Daniel and De Petris, Marco and Ettori, Stefano and Haggar, Roan and Li, Qingyang and Wang, Yang and Yang, Xiaohu and Borgani, Stefano and Dolag, Klaus and Zu, Ying and Kuchner, Ulrike and Cañas, Rodrigo and Ferragamo, Antonio and Gianfagna, Giulia},
        title = {The Three Hundred project: The GIZMO-SIMBA run},
      journal = {MNRAS},
         year = 2022,
       volume = {514},
       number = {1},
        pages = {977--996},
          doi = {10.1093/mnras/stac1402}
}

@article{cui_three_2018,
       author = {Cui, Weiguang and Knebe, Alexander and Yepes, Gustavo and Pearce, Frazer and Power, Chris and Dave, Romeel and Arth, Alexander and Borgani, Stefano and Dolag, Klaus and Elahi, Pascal and Mostoghiu, Robert and Murante, Giuseppe and Rasia, Elena and Stoppacher, Doris and Vega-Ferrero, Jesus and Wang, Yang and Yang, Xiaohu and Benson, Andrew and Cora, Sofía A and Croton, Darren J and Sinha, Manodeep and Stevens, Adam R H and Vega-Martínez, Cristian A and Arthur, Jake and Baldi, Anna S and Cañas, Rodrigo and Cialone, Giammarco and Cunnama, Daniel and De Petris, Marco and Durando, Giacomo and Ettori, Stefano and Gottlöber, Stefan and Nuza, Sebastián E and Old, Lyndsay J and Pilipenko, Sergey and Sorce, Jenny G and Welker, Charlotte},
        title = {The Three Hundred project: a large catalogue of theoretically modelled galaxy clusters for cosmological and astrophysical applications},
      journal = {MNRAS},
         year = 2018,
       volume = {480},
       number = {3},
        pages = {2898--2915},
          doi = {10.1093/mnras/sty2111}
}

@article{diaferio_mass_1999,
       author = {{Diaferio}, Antonaldo},
        title = {Mass estimation in the outer regions of galaxy clusters},
      journal = {MNRAS},
         year = 1999,
          doi = {10.1046/j.1365-8711.1999.02864.x}
}

@article{serra_measuring_2010,
       author = {{Serra}, Ana Laura and {Diaferio}, Antonaldo and {Murante}, Giuseppe and {Borgani}, Stefano},
        title = {Measuring the escape velocity and mass profiles of galaxy clusters beyond their virial radius},
      journal = {MNRAS},
         year = 2010,
       pages = {no--no}
}

@article{gifford_systematic_2013,
       author = {{Gifford}, Daniel and {Miller}, Christopher and {Kern}, Nicholas},
        title = {A systematic analysis of caustic methods for galaxy cluster masses},
      journal = {ApJ},
         year = 2013,
       volume = {773},
       number = {2},
        pages = {116},
          doi = {10.1088/0004-637X/773/2/116}
}

@article{geller_measuring_2013,
       author = {{Geller}, Margaret J. and {Diaferio}, Antonaldo and {Rines}, Kenneth J. and {Serra}, Ana Laura},
        title = {Measuring the mass distribution in galaxy clusters},
      journal = {ApJ},
         year = 2013,
       volume = {764},
       number = {1},
        pages = {58},
          doi = {10.1088/0004-637X/764/1/58}
}

@article{mamon_dark_2005,
    title = {Dark matter in elliptical galaxies - II. Estimating the mass within the virial radius},
    volume = {363},
    number = {3},
    pages = {705--722},
    year = {2005},
    journal = {MNRAS},
    author = {Mamon, Gary A. and Łokas, Ewa L.},
    doi = {10.1111/j.1365-2966.2005.09400.x}
}

@article{blumenthal_formation_1984,
    title = {Formation of galaxies and large-scale structure with cold dark matter},
    volume = {311},
    number = {5986},
    pages = {517--525},
    year = {1984},
    journal = {Nature},
    author = {Blumenthal, George R. and Faber, Sandra M. and Primack, Joel R. and Rees, Martin J.},
    doi = {10.1038/311517a0}
}

@article{milgrom_mond--theoretical_2002,
    title = {{MOND}--theoretical aspects},
    volume = {46},
    number = {12},
    pages = {741--753},
    year = {2002},
    journal = {New Astron. Rev.},
    author = {Milgrom, Mordehai},
    doi = {10.1016/S1387-6473(02)00243-9}
}

@article{klypin_multidark_2016,
    title = {MultiDark simulations: the story of dark matter halo concentrations and density profiles},
    volume = {457},
    number = {4},
    pages = {4340--4359},
    year = {2016},
    journal = {MNRAS},
    author = {Klypin, Anatoly and Yepes, Gustavo and Gottlöber, Stefan and Prada, Francisco and Heß, Steffen},
    doi = {10.1093/mnras/stw248}
}

@article{planck_collaboration_planck_2016,
    title = {Planck 2015 results. XIII. Cosmological parameters},
    volume = {594},
    pages = {A13},
    year = {2016},
    journal = {A\&A},
    author = {{Planck Collaboration} and Ade, P. A. R. and Aghanim, N. and Arnaud, M. and Ashdown, M. and Aumont, J. and Baccigalupi, C. and Banday, A. J. and Barreiro, R. B. and Bartlett, J. G. and Bartolo, N. and Battaner, E. and Battye, R. and Benabed, K. and Benoît, A. and Benoit-Lévy, A. and Bernard, J.-P. and Bersanelli, M. and Bielewicz, P. and Bock, J. J. and Bonaldi, A. and Bonavera, L. and Bond, J. R. and Borrill, J. and Bouchet, F. R. and Boulanger, F. and Bucher, M. and Burigana, C. and Butler, R. C. and Calabrese, E. and Cardoso, J.-F. and Catalano, A. and Challinor, A. and Chamballu, A. and Chary, R.-R. and Chiang, H. C. and Chluba, J. and Christensen, P. R. and Church, S. and Clements, D. L. and Colombi, S. and Colombo, L. P. L. and Combet, C. and Coulais, A. and Crill, B. P. and Curto, A. and Cuttaia, F. and Danese, L. and Davies, R. D. and Davis, R. J. and De Bernardis, P. and De Rosa, A. and De Zotti, G. and Delabrouille, J. and Désert, F.-X. and Di Valentino, E. and Dickinson, C. and Diego, J. M. and Dolag, K. and Dole, H. and Donzelli, S. and Doré, O. and Douspis, M. and Ducout, A. and Dunkley, J. and Dupac, X. and Efstathiou, G. and Elsner, F. and Enßlin, T. A. and Eriksen, H. K. and Farhang, M. and Fergusson, J. and Finelli, F. and Forni, O. and Frailis, M. and Fraisse, A. A. and Franceschi, E. and Frejsel, A. and Galeotta, S. and Galli, S. and Ganga, K. and Gauthier, C. and Gerbino, M. and Ghosh, T. and Giard, M. and Giraud-Héraud, Y. and Giusarma, E. and Gjerløw, E. and González-Nuevo, J. and Górski, K. M. and Gratton, S. and Gregorio, A. and Gruppuso, A. and Gudmundsson, J. E. and Hamann, J. and Hansen, F. K. and Hanson, D. and Harrison, D. L. and Helou, G. and Henrot-Versillé, S. and Hernández-Monteagudo, C. and Herranz, D. and Hildebrandt, S. R. and Hivon, E. and Hobson, M. and Holmes, W. A. and Hornstrup, A. and Hovest, W. and Huang, Z. and Huffenberger, K. M. and Hurier, G. and Jaffe, A. H. and Jaffe, T. R. and Jones, W. C. and Juvela, M. and Keihänen, E. and Keskitalo, R. and Kisner, T. S. and Kneissl, R. and Knoche, J. and Knox, L. and Kunz, M. and Kurki-Suonio, H. and Lagache, G. and Lähteenmäki, A. and Lamarre, J.-M. and Lasenby, A. and Lattanzi, M. and Lawrence, C. R. and Leahy, J. P. and Leonardi, R. and Lesgourgues, J. and Levrier, F. and Lewis, A. and Liguori, M. and Lilje, P. B. and Linden-Vørnle, M. and López-Caniego, M. and Lubin, P. M. and Macías-Pérez, J. F. and Maggio, G. and Maino, D. and Mandolesi, N. and Mangilli, A. and Marchini, A. and Maris, M. and Martin, P. G. and Martinelli, M. and Martínez-González, E. and Masi, S. and Matarrese, S. and McGehee, P. and Meinhold, P. R. and Melchiorri, A. and Melin, J.-B. and Mendes, L. and Mennella, A. and Migliaccio, M. and Millea, M. and Mitra, S. and Miville-Deschênes, M.-A. and Moneti, A. and Montier, L. and Morgante, G. and Mortlock, D. and Moss, A. and Munshi, D. and Murphy, J. A. and Naselsky, P. and Nati, F. and Natoli, P. and Netterfield, C. B. and Nørgaard-Nielsen, H. U. and Noviello, F. and Novikov, D. and Novikov, I. and Oxborrow, C. A. and Paci, F. and Pagano, L. and Pajot, F. and Paladini, R. and Paoletti, D. and Partridge, B. and Pasian, F. and Patanchon, G. and Pearson, T. J. and Perdereau, O. and Perotto, L. and Perrotta, F. and Pettorino, V. and Piacentini, F. and Piat, M. and Pierpaoli, E. and Pietrobon, D. and Plaszczynski, S. and Pointecouteau, E. and Polenta, G. and Popa, L. and Pratt, G. W. and Prézeau, G. and Prunet, S. and Puget, J.-L. and Rachen, J. P. and Reach, W. T. and Rebolo, R. and Reinecke, M. and Remazeilles, M. and Renault, C. and Renzi, A. and Ristorcelli, I. and Rocha, G. and Rosset, C. and Rossetti, M. and Roudier, G. and Rouillé d’Orfeuil, B. and Rowan-Robinson, M. and Rubiño-Martín, J. A. and Rusholme, B. and Said, N. and Salvatelli, V. and Salvati, L. and Sandri, M. and Santos, D. and Savelainen, M. and Savini, G. and Scott, D. and Seiffert, M. D. and Serra, P. and Shellard, E. P. S. and Spencer, L. D. and Spinelli, M. and Stolyarov, V. and Stompor, R. and Sudiwala, R. and Sunyaev, R. and Sutton, D. and Suur-Uski, A.-S. and Sygnet, J.-F. and Tauber, J. A. and Terenzi, L. and Toffolatti, L. and Tomasi, M. and Tristram, M. and Trombetti, T. and Tucci, M. and Tuovinen, J. and Türler, M. and Umana, G. and Valenziano, L. and Valiviita, J. and Van Tent, F. and Vielva, P. and Villa, F. and Wade, L. A. and Wandelt, B. D. and Wehus, I. K. and White, M. and White, S. D. M. and Wilkinson, A. and Yvon, D. and Zacchei, A. and Zonca, A.},
    doi = {10.1051/0004-6361/201525830}
}

@article{foex_comparison_2017,
    title = {Comparison of hydrostatic and dynamical masses of distant X-ray luminous galaxy clusters},
    volume = {606},
    pages = {A122},
    year = {2017},
    journal = {A\&A},
    author = {Foëx, G. and Böhringer, H. and Chon, G.},
    doi = {10.1051/0004-6361/201731104}
}

@article{diaferio_infall_1997,
    title = {Infall regions of galaxy clusters},
    volume = {481},
    number = {2},
    pages = {633--643},
    year = {1997},
    journal = {ApJ},
    author = {Diaferio, Antonaldo and Geller, Margaret J.},
    doi = {10.1086/304075}
}

@article{ansarifard_three_2020,
    title = {The Three Hundred Project: Correcting for the hydrostatic-equilibrium mass bias in X-ray and SZ surveys},
    volume = {634},
    pages = {A113},
    year = {2020},
    journal = {A\&A},
    author = {Ansarifard, S. and Rasia, E. and Biffi, V. and Borgani, S. and Cui, W. and De Petris, M. and Dolag, K. and Ettori, S. and Movahed, S. M. S. and Murante, G. and Yepes, G.},
    doi = {10.1051/0004-6361/201936742}
}

@article{geller_mass_1999,
    title = {The mass profile of the Coma galaxy cluster},
    volume = {517},
    number = {1},
    pages = {L23--L26},
    year = {1999},
    journal = {ApJL},
    author = {Geller, M. J. and Diaferio, A. and Kurtz, M. J.},
    doi = {10.1086/312024}
}

@article{haggar_thethreehundred_2020,
    title = {The Three Hundred project: backsplash galaxies in simulations of clusters},
    volume = {492},
    number = {4},
    pages = {6074--6085},
    year = {2020},
    journal = {MNRAS},
    author = {Haggar, Roan and Gray, Meghan E and Pearce, Frazer R and Knebe, Alexander and Cui, Weiguang and Mostoghiu, Robert and Yepes, Gustavo},
    doi = {10.1093/mnras/staa273}
}

@article{knebe_three_2020,
    title = {The Three Hundred project: shapes and radial alignment of satellite, infalling, and backsplash galaxies},
    volume = {495},
    number = {3},
    pages = {3002--3013},
    year = {2020},
    journal = {MNRAS},
    author = {Knebe, Alexander and Gámez-Marín, Matías and Pearce, Frazer R and Cui, Weiguang and Hoffmann, Kai and De Petris, Marco and Power, Chris and Haggar, Roan and Mostoghiu, Robert},
    doi = {10.1093/mnras/staa1407}
}

@article{pizzardo_illustristng_2023,
    title = {An IllustrisTNG view of the caustic technique for galaxy cluster mass estimation},
    volume = {675},
    pages = {A56},
    year = {2023},
    journal = {A\&A},
    author = {Pizzardo, Michele and Geller, Margaret J. and Kenyon, Scott J. and Damjanov, Ivana and Diaferio, Antonaldo},
    doi = {10.1051/0004-6361/202346545}
}

@article{mcgaugh_tale_2015,
    title = {A tale of two paradigms: the mutual incommensurability of ΛCDM and MOND},
    volume = {93},
    number = {2},
    pages = {250--259},
    year = {2015},
    journal = {Can. J. Phys.},
    author = {McGaugh, S. S.},
    doi = {10.1139/cjp-2014-0203}
}

@article{pizzardo_mass_2021,
    title = {Mass accretion rates of clusters of galaxies: CIRS and HeCS},
    volume = {646},
    journal = {A\&A},
    author = {Pizzardo, M. and Di Gioia, S. and Diaferio, A. and De Boni, C. and Serra, A. L. and Geller, M. J. and Sohn, J. and Rines, K. and Baldi, M.},
    year = {2021},
    pages = {A105},
    doi = {10.1051/0004-6361/202038481}
}

@article{maughan_hydrostatic_2016,
    title = {Hydrostatic and caustic mass profiles of galaxy clusters},
    volume = {461},
    journal = {MNRAS},
    author = {Maughan, Ben J. and Giles, Paul A. and Rines, Kenneth J. and Diaferio, Antonaldo and Geller, Margaret J. and Van Der Pyl, Nina and Bonamente, Massimiliano},
    year = {2016},
    pages = {4182--4191},
    doi = {10.1093/mnras/stw1610}
}

@article{rines_cirs_2006,
    title = {CIRS: Cluster Infall Regions in the Sloan Digital Sky Survey. I. Infall Patterns and Mass Profiles},
    volume = {132},
    journal = {AJ},
    author = {Rines, Kenneth and Diaferio, Antonaldo},
    year = {2006},
    pages = {1275--1297},
    doi = {10.1086/506017}
}

@article{rines_measuring_2013,
    title = {Measuring the Ultimate Halo Mass of Galaxy Clusters: Redshifts and Mass Profiles from the HeCS},
    volume = {767},
    journal = {ApJ},
    author = {Rines, Kenneth and Geller, Margaret J. and Diaferio, Antonaldo and Kurtz, Michael J.},
    year = {2013},
    pages = {15},
    doi = {10.1088/0004-637X/767/1/15}
}

@article{umetsu_model-free_2013,
    title = {Model-free multi-probe lensing reconstruction of cluster mass profiles},
    volume = {769},
    journal = {ApJ},
    author = {Umetsu, Keiichi},
    year = {2013},
    pages = {13},
    doi = {10.1088/0004-637X/769/1/13}
}

@article{umetsu_cluster_2011,
    title = {Cluster mass profiles from a Bayesian analysis of weak-lensing distortion and magnification measurements},
    volume = {729},
    journal = {ApJ},
    author = {Umetsu, Keiichi and Broadhurst, Tom and Zitrin, Adi and Medezinski, Elinor and Hsu, Li-Yen},
    year = {2011},
    pages = {127},
    doi = {10.1088/0004-637X/729/2/127}
}

@article{armitage_cluster-eagle_2019,
    title = {The Cluster-EAGLE project: a comparison of dynamical mass estimators using simulated clusters},
    volume = {482},
    journal = {MNRAS},
    author = {Armitage, Thomas J and Kay, Scott T and Barnes, David J and Bahé, Yannick M and Dalla Vecchia, Claudio},
    year = {2019},
    pages = {3308--3325},
    doi = {10.1093/mnras/sty2921}
}

@article{sembolini_music_2013,
    title = {The MUSIC of galaxy clusters – I. Baryon properties and scaling relations of the thermal Sunyaev–Zel'dovich effect},
    journal = {MNRAS},
    volume = {429},
    pages = {323--343},
    year = {2013},
    doi = {10.1093/mnras/sts339},
    author = {Sembolini, Federico and Yepes, Gustavo and De Petris, Marco and Gottlöber, Stefan and Lamagna, Luca and Comis, Barbara}
}

@article{rasia_cool_2015,
    title = {Cool core clusters from cosmological simulations},
    journal = {ApJL},
    volume = {813},
    pages = {L17},
    year = {2015},
    doi = {10.1088/2041-8205/813/1/L17},
    author = {Rasia, E. and Borgani, S. and Murante, G. and Planelles, S. and Beck, A. M. and Biffi, V. and Ragone-Figueroa, C. and Granato, G. L. and Steinborn, L. K. and Dolag, K.}
}

@article{rines_infall_2000,
    title = {The infall region of Abell 576},
    journal = {AJ},
    volume = {120},
    pages = {2338--2354},
    year = {2000},
    author = {Rines, Kenneth and Geller, Margaret J. and Diaferio, Antonaldo and Mohr, Joseph J. and Wegner, Gary A.}
}

@article{rines_mass_2002,
    title = {Mass profile of the infall region of the Abell 2199 supercluster},
    journal = {AJ},
    volume = {124},
    pages = {1266--1282},
    year = {2002},
    author = {Rines, K. and Geller, M. J. and Diaferio, A. and Mahdavi, A. and Mohr, J. J. and Wegner, G.}
}

@article{bakels_pre-processing_2021,
    title = {Pre-processing, group accretion, and orbital trajectories of subhaloes},
    journal = {MNRAS},
    volume = {501},
    pages = {5948--5963},
    year = {2021},
    doi = {10.1093/mnras/staa3979},
    author = {Bakels, Lucie and Ludlow, Aaron D and Power, Chris}
}

@article{diaferio_caustic_2005,
    title = {Caustic and weak-lensing estimators of galaxy cluster masses},
    journal = {ApJ},
    volume = {628},
    pages = {L97--L100},
    year = {2005},
    doi = {10.1086/432880},
    author = {Diaferio, Antonaldo and Geller, Margaret J. and Rines, Kenneth J.}
}

@article{ludlow_dynamical_2012,
    title = {The dynamical state and mass–concentration relation of galaxy clusters},
    journal = {MNRAS},
    volume = {427},
    pages = {1322--1328},
    year = {2012},
    author = {Ludlow, Aaron D. and Navarro, Julio F. and Li, Ming and Angulo, Raul E. and Boylan-Kolchin, Michael and Bett, Philip E.}
}

@article{nelson_weighing_2014,
    title = {Weighing galaxy clusters with gas: origin of hydrostatic mass bias},
    journal = {ApJ},
    volume = {782},
    pages = {107},
    year = {2014},
    doi = {10.1088/0004-637X/782/2/107},
    author = {Nelson, Kaylea and Lau, Erwin T. and Nagai, Daisuke and Rudd, Douglas H. and Yu, Liang}
}

@article{rasia_lensing_2012,
    title = {Lensing and X-ray mass estimates of clusters},
    journal = {New J Phys},
    volume = {14},
    pages = {055018},
    year = {2012},
    doi = {10.1088/1367-2630/14/5/055018},
    author = {Rasia, E and Meneghetti, M and Martino, R and Borgani, S and Bonafede, A and Dolag, K and Ettori, S and Fabjan, D and Giocoli, C and Mazzotta, P and Merten, J and Radovich, M and Tornatore, L}
}

@article{umetsu_weak-lensing_2020,
    title = {Weak-lensing analysis of XXL clusters with HSC data},
    journal = {ApJ},
    volume = {890},
    pages = {148},
    year = {2020},
    doi = {10.3847/1538-4357/ab6bca},
    author = {Umetsu, Keiichi and Sereno, Mauro and Lieu, Maggie and Miyatake, Hironao and Medezinski, Elinor and Nishizawa, Atsushi J. and Giles, Paul and Gastaldello, Fabio and McCarthy, Ian G. and Kilbinger, Martin and Birkinshaw, Mark and Ettori, Stefano and Okabe, Nobuhiro and Chiu, I-Non and Coupon, Jean and Eckert, Dominique and Fujita, Yutaka and Higuchi, Yuichi and Koulouridis, Elias and Maughan, Ben and Miyazaki, Satoshi and Oguri, Masamune and Pacaud, Florian and Pierre, Marguerite and Rapetti, David and Smith, Graham P.}
}

@article{merritt_distribution_1987,
    title = {The distribution of dark matter in the Coma cluster},
    journal = {ApJ},
    volume = {313},
    pages = {121},
    year = {1987},
    author = {Merritt, David}
}

@article{mcclintock_dark_2019,
    title = {Dark Energy Survey Year 1 weak lensing mass calibration of redMaPPer clusters},
    journal = {MNRAS},
    volume = {482},
    pages = {1352--1378},
    year = {2019},
    doi = {10.1093/mnras/sty2711},
    author = {McClintock, T and Varga, T N and Gruen, D and Rozo, E and Rykoff, E S and Shin, T and Melchior, P and DeRose, J and Seitz, S and Dietrich, J P and Sheldon, E and Zhang, Y and von der Linden, A and Jeltema, T and Mantz, A B and Romer, A K and Allen, S and Becker, M R and Bermeo, A and Bhargava, S and Costanzi, M and Everett, S and Farahi, A and Hamaus, N and Hartley, W G and Hollowood, D L and Hoyle, B and Israel, H and Li, P and MacCrann, N and Morris, G and Palmese, A and Plazas, A A and Pollina, G and Rau, M M and Simet, M and Soares-Santos, M and Troxel, M A and Vergara Cervantes, C and Wechsler, R H and Zuntz, J and Abbott, T M C and Abdalla, F B and Allam, S and Annis, J and Avila, S and Bridle, S L and Brooks, D and Burke, D L and Carnero Rosell, A and Carrasco Kind, M and Carretero, J and Castander, F J and Crocce, M and Cunha, C E and D’Andrea, C B and da Costa, L N and Davis, C and De Vicente, J and Diehl, H T and Doel, P and Drlica-Wagner, A and Evrard, A E and Flaugher, B and Fosalba, P and Frieman, J and García-Bellido, J and Gaztanaga, E and Gerdes, D W and Giannantonio, T and Gruendl, R A and Gutierrez, G and Honscheid, K and James, D J and Kirk, D and Krause, E and Kuehn, K and Lahav, O and Li, T S and Lima, M and March, M and Marshall, J L and Menanteau, F and Miquel, R and Mohr, J J and Nord, B and Ogando, R L C and Roodman, A and Sanchez, E and Scarpine, V and Schindler, R and Sevilla-Noarbe, I and Smith, M and Smith, R C and Sobreira, F and Suchyta, E and Swanson, M E C and Tarle, G and Tucker, D L and Vikram, V and Walker, A R and Weller, J and {DES Collaboration}}
}

@article{capalbo_three_2021,
    title = {The Three Hundred project: cluster morphology via Zernike polynomials},
    journal = {MNRAS},
    volume = {503},
    pages = {6155--6169},
    year = {2021},
    doi = {10.1093/mnras/staa3900},
    author = {Capalbo, Valentina and De Petris, Marco and De Luca, Federico and Cui, Weiguang and Yepes, Gustavo and Knebe, Alexander and Rasia, Elena}
}

@article{contreras-santos_three_2022,
    title = {Galaxy cluster mergers and BCG evolution in The Three Hundred project},
    journal = {MNRAS},
    volume = {511},
    pages = {2897--2913},
    year = {2022},
    doi = {10.1093/mnras/stac275},
    author = {Contreras-Santos, Ana and Knebe, Alexander and Pearce, Frazer and Haggar, Roan and Gray, Meghan and Cui, Weiguang and Yepes, Gustavo and De Petris, Marco and De Luca, Federico and Power, Chris and Mostoghiu, Robert and Nuza, Sebastián E and Hoeft, Matthias}
}

@article{deluca_three_2021,
    title = {Dynamical state of galaxy clusters in The Three Hundred project},
    journal = {MNRAS},
    volume = {504},
    pages = {5383--5400},
    year = {2021},
    doi = {10.1093/mnras/stab1073},
    author = {De Luca, Federico and De Petris, Marco and Yepes, Gustavo and Cui, Weiguang and Knebe, Alexander and Rasia, Elena}
}

@article{gianfagna_hydrostatic_2022,
    title = {Hydrostatic mass bias in The Three Hundred clusters},
    journal = {EPJ Conf},
    volume = {257},
    pages = {00020},
    year = {2022},
    doi = {10.1051/epjconf/202225700020},
    author = {Gianfagna, Giulia and Rasia, Elena and Cui, Weiguang and De Petris, Marco and Yepes, Gustavo}
}

@article{wang_three_2018,
    title = {The Three Hundred Project: The Influence of Environment on Simulated Galaxy Properties},
    volume = {868},
    doi = {10.3847/1538-4357/aae52e},
    journal = {ApJ},
    author = {Wang, Yang and Pearce, Frazer and Knebe, Alexander and Yepes, Gustavo and Cui, Weiguang and Power, Chris and Arth, Alexander and Gottlöber, Stefan and De Petris, Marco and Brown, Shaun and Feng, Longlong},
    year = {2018},
    pages = {130}
}

@article{zhang_span_2022,
    title = {The Three Hundred: cluster dynamical states and relaxation period},
    volume = {516},
    doi = {10.1093/mnras/stac2171},
    journal = {MNRAS},
    author = {Zhang, Bowei and Cui, Weiguang and Wang, Yuhuan and Dave, Romeel and De Petris, Marco},
    year = {2022},
    pages = {26--38}
}

@article{Anbajagane_shocks_2022,
    author = {Anbajagane, D and Chang, C and Jain, B and Adhikari, S and Baxter, E J and Benson, B A and Bleem, L E and Bocquet, S and Calzadilla, M S and Carlstrom, J E and Chang, C L and Chown, R and Crawford, T M and Crites, A T and Cui, W and de Haan, T and Mascolo, L Di and Dobbs, M A and Everett, W B and George, E M and Grandis, S and Halverson, N W and Holder, G P and Holzapfel, W L and Hrubes, J D and Lee, A T and Luong-Van, D and McDonald, M A and McMahon, J J and Meyer, S S and Millea, M and Mocanu, L M and Mohr, J J and Natoli, T and Omori, Y and Padin, S and Pryke, C and Reichardt, C L and Ruhl, J E and Saro, A and Schaffer, K K and Shirokoff, E and Staniszewski, Z and Stark, A A and Vieira, J D and Williamson, R},
    title = {Shocks in the stacked Sunyaev-Zel'dovich profiles of clusters II: Measurements from SPT-SZ + Planck Compton-y map},
    journal = {MNRAS},
    year = {2022},
    volume = {514},
    number = {2},
    pages = {1645--1663},
    doi = {10.1093/mnras/stac1376}
}

@article{arthur_span_2019,
    title = {TheThreeHundred Project: ram pressure and gas content of haloes and subhaloes in the phase-space plane},
    volume = {484},
    doi = {10.1093/mnras/stz212},
    journal = {MNRAS},
    author = {Arthur, Jake and Pearce, Frazer R and Gray, Meghan E and Knebe, Alexander and Cui, Weiguang and Elahi, Pascal J and Power, Chris and Yepes, Gustavo and De Petris, Marco},
    year = {2019},
    pages = {3968--3983}
}

@article{baxter_shocks_2021,
    title = {Shocks in the stacked Sunyaev–Zel'dovich profiles of clusters – I. Analysis with the Three Hundred simulations},
    volume = {508},
    doi = {10.1093/mnras/stab2720},
    journal = {MNRAS},
    author = {Baxter, Eric J and Adhikari, Susmita and Vega-Ferrero, Jesús and Cui, Weiguang and Chang, Chihway and Jain, Bhuvnesh and Knebe, Alexander},
    year = {2021},
    pages = {1777--1787}
}

@article{becker_accuracy_2011,
    title = {On the accuracy of weak-lensing cluster mass reconstructions},
    volume = {740},
    doi = {10.1088/0004-637X/740/1/25},
    journal = {ApJ},
    author = {Becker, Matthew R. and Kravtsov, Andrey V.},
    year = {2011},
    pages = {25}
}

@article{corless_new_2009,
    title = {A new look at massive clusters: weak lensing constraints on the triaxial dark matter haloes of A1689, A1835 and A2204},
    volume = {393},
    doi = {10.1111/j.1365-2966.2008.14294.x},
    journal = {MNRAS},
    author = {Corless, Virginia L. and King, Lindsay J. and Clowe, Douglas},
    year = {2009},
    pages = {1235--1254}
}

@article{feroz_weak_2012,
    title = {Weak lensing by triaxial galaxy clusters},
    volume = {420},
    doi = {10.1111/j.1365-2966.2011.20070.x},
    journal = {MNRAS},
    author = {Feroz, F. and Hobson, M. P.},
    year = {2012},
    pages = {596--603}
}

@article{hoekstra_how_2003,
    title = {How well can we determine cluster mass profiles from weak lensing?},
    volume = {339},
    doi = {10.1046/j.1365-8711.2003.06264.x},
    journal = {MNRAS},
    author = {Hoekstra, Henk},
    year = {2003},
    pages = {1155--1162}
}

@article{kotecha_cosmic_2022,
    title = {Cosmic filaments delay quenching inside clusters},
    volume = {512},
    doi = {10.1093/mnras/stac300},
    journal = {MNRAS},
    author = {Kotecha, Sachin and Welker, Charlotte and Zhou, Zihan and Wadsley, James and Kraljic, Katarina and Sorce, Jenny and Rasia, Elena and Roberts, Ian and Gray, Meghan and Yepes, Gustavo and Cui, Weiguang},
    year = {2022},
    pages = {926--944}
}

@article{kuchner_cosmic_2021,
    title = {Cosmic filaments in galaxy cluster outskirts: quantifying finding filaments in redshift space},
    volume = {503},
    doi = {10.1093/mnras/stab567},
    journal = {MNRAS},
    author = {Kuchner, Ulrike and Aragón-Salamanca, Alfonso and Rost, Agustín and Pearce, Frazer R and Gray, Meghan E and Cui, Weiguang and Knebe, Alexander},
    year = {2021},
    pages = {2065--2076}
}

@article{li_three_2020,
       author = {Li, Qingyang and Cui, Weiguang and Yang, Xiaohu and Rasia, Elena and Dav{\'e}, Romeel and De Petris, Marco and Knebe, Alexander and Peacock, John A. and Pearce, Frazer and Yepes, Gustavo},
        title = {The Three Hundred project: the stellar and gas profiles},
      journal = {MNRAS},
         year = {2020},
       volume = {495},
       number = {3},
        pages = {2930-2948},
          doi = {10.1093/mnras/staa1385}
}

@article{mostoghiu_three_2019,
    title = {The Three Hundred Project: The evolution of galaxy cluster density profiles},
    volume = {483},
    doi = {10.1093/mnras/sty3306},
    journal = {MNRAS},
    author = {Mostoghiu, Robert and Knebe, Alexander and Cui, Weiguang and Pearce, Frazer R and Yepes, Gustavo and Power, Chris and Dave, Romeel and Arth, Alexander},
    year = {2019},
    pages = {3390--3403}
}

@article{jin_wide-field_2024,
    title = {The wide-field, multiplexed, spectroscopic facility WEAVE: Survey design, overview, and simulated implementation},
    volume = {530},
    doi = {10.1093/mnras/stad557},
    journal = {MNRAS},
    author = {Jin, Shoko and Trager, Scott C and Dalton, Gavin B and Aguerri, J Alfonso L and Drew, J E and Falcón-Barroso, Jesús and Gänsicke, Boris T and Hill, Vanessa and Iovino, Angela and Pieri, Matthew M and Poggianti, Bianca M and Smith, D J B and Vallenari, Antonella and Abrams, Don Carlos and Aguado, David S and Antoja, Teresa and Aragón-Salamanca, Alfonso and Ascasibar, Yago and Babusiaux, Carine and Balcells, Marc and Barrena, R and Battaglia, Giuseppina and Belokurov, Vasily and Bensby, Thomas and Bonifacio, Piercarlo and Bragaglia, Angela and Carrasco, Esperanza and Carrera, Ricardo and Cornwell, Daniel J and Domínguez-Palmero, Lilian and Duncan, Kenneth J and Famaey, Benoit and Fariña, Cecilia and Gonzalez, Oscar A and Guest, Steve and Hatch, Nina A and Hess, Kelley M and Hoskin, Matthew J and Irwin, Mike and Knapen, Johan H and Koposov, Sergey E and Kuchner, Ulrike and Laigle, Clotilde and Lewis, Jim and Longhetti, Marcella and Lucatello, Sara and Méndez-Abreu, Jairo and Mercurio, Amata and Molaeinezhad, Alireza and Monguió, Maria and Morrison, Sean and Murphy, David N A and Peralta de Arriba, Luis and Pérez, Isabel and Pérez-Ràfols, Ignasi and Picó, Sergio and Raddi, Roberto and Romero-Gómez, Mercè and Royer, Frédéric and Siebert, Arnaud and Seabroke, George M and Som, Debopam and Terrett, David and Thomas, Guillaume and Wesson, Roger and Worley, C Clare and Alfaro, Emilio J and Allende Prieto, Carlos and Alonso-Santiago, Javier and Amos, Nicholas J and Ashley, Richard P and Balaguer-Núñez, Lola and Balbinot, Eduardo and Bellazzini, Michele and Benn, Chris R and Berlanas, Sara R and Bernard, Edouard J and Best, Philip and Bettoni, Daniela and Bianco, Andrea and Bishop, Georgia and Blomqvist, Michael and Boeche, Corrado and Bolzonella, Micol and Bonoli, Silvia and Bosma, Albert and Britavskiy, Nikolay and Busarello, Gianni and Caffau, Elisabetta and Cantat-Gaudin, Tristan and Castro-Ginard, Alfred and Couto, Guilherme and Carbajo-Hijarrubia, Juan and Carter, David and Casamiquela, Laia and Conrado, Ana M and Corcho-Caballero, Pablo and Costantin, Luca and Deason, Alis and de Burgos, Abel and De Grandi, Sabrina and Di Matteo, Paola and Domínguez-Gómez, Jesús and Dorda, Ricardo and Drake, Alyssa and Dutta, Rajeshwari and Erkal, Denis and Feltzing, Sofia and Ferré-Mateu, Anna and Feuillet, Diane and Figueras, Francesca and Fossati, Matteo and Franciosini, Elena and Frasca, Antonio and Fumagalli, Michele and Gallazzi, Anna and García-Benito, Rubén and Gentile Fusillo, Nicola and Gebran, Marwan and Gilbert, James and Gledhill, T M and González Delgado, Rosa M and Greimel, Robert and Guarcello, Mario Giuseppe and Guerra, Jose and Gullieuszik, Marco and Haines, Christopher P and Hardcastle, Martin J and Harris, Amy and Haywood, Misha and Helmi, Amina and Hernandez, Nauzet and Herrero, Artemio and Hughes, Sarah and Iršič, Vid and Jablonka, Pascale and Jarvis, Matt J and Jordi, Carme and Kondapally, Rohit and Kordopatis, Georges and Krogager, Jens-Kristian and La Barbera, Francesco and Lam, Man I and Larsen, Søren S and Lemasle, Bertrand and Lewis, Ian J and Lhomé, Emilie and Lind, Karin and Lodi, Marcello and Longobardi, Alessia and Lonoce, Ilaria and Magrini, Laura and Maíz Apellániz, Jesús and Marchal, Olivier and Marco, Amparo and Martin, Nicolas F and Matsuno, Tadafumi and Maurogordato, Sophie and Merluzzi, Paola and Miralda-Escudé, Jordi and Molinari, Emilio and Monari, Giacomo and Morelli, Lorenzo and Mottram, Christopher J and Naylor, Tim and Negueruela, Ignacio and Oñorbe, Jose and Pancino, Elena and Peirani, Sébastien and Peletier, Reynier F and Pozzetti, Lucia and Rainer, Monica and Ramos, Pau and Read, Shaun C and Rossi, Elena Maria and Röttgering, Huub J A and Rubiño-Martín, Jose Alberto and Sabater, Jose and San Juan, José and Sanna, Nicoletta and Schallig, Ellen and Schiavon, Ricardo P and Schultheis, Mathias and Serra, Paolo and Shimwell, Timothy W and Simón-Díaz, Sergio and Smith, Russell J and Sordo, Rosanna and Sorini, Daniele and Soubiran, Caroline and Starkenburg, Else and Steele, Iain A and Stott, John and Stuik, Remko and Tolstoy, Eline and Tortora, Crescenzo and Tsantaki, Maria and Van der Swaelmen, Mathieu and van Weeren, Reinout J and Vergani, Daniela and Verheijen, Marc A W and Verro, Kristiina and Vink, Jorick S and Vioque, Miguel and Walcher, C Jakob and Walton, Nicholas A and Wegg, Christopher and Weijmans, Anne-Marie and Williams, Wendy L and Wilson, Andrew J and Wright, Nicholas J and Xylakis-Dornbusch, Theodora and Youakim, Kris and Zibetti, Stefano and Zurita, Cristina},
    year = {2024},
    pages = {2688--2730}
}

@article{4most_2019_e_ROSITA,
       author = {{Finoguenov}, A. and {Merloni}, A. and {Comparat}, J. and {Nandra}, K. and {Salvato}, M. and {Tempel}, E. and {Raichoor}, A. and {Richard}, J. and {Kneib}, J.-P. and {Pillepich}, A. and {Sahl{\'e}n}, M. and {Popesso}, P. and {Norberg}, P. and {McMahon}, R. and {4MOST Collaboration}},
        title = "{4MOST Consortium Survey 5: eROSITA Galaxy Cluster Redshift Survey}",
      journal = {The Messenger},
         year = 2019,
       volume = {175},
        pages = {39-41},
          doi = {10.18727/0722-6691/5124}
}

@article{rines_hecs-sz_2016,
  title = {HeCS-SZ: The Hectospec Survey of Sunyaev–Zel’dovich-selected clusters},
  volume = {819},
  doi = {10.3847/0004-637X/819/1/63},
  journal = {ApJ},
  author = {Rines, Kenneth J. and Geller, Margaret J. and Diaferio, Antonaldo and Hwang, Ho Seong},
  year = {2016},
  pages = {63}
}

@book{sarazin_1988,
  title = {X-ray Emission from Clusters of Galaxies},
  author = {Sarazin, Craig L.},
  year = {1988},
  publisher = {Cambridge University Press}
}

@article{the_and_white_1986,
  title = {The mass of the Coma cluster},
  journal = {AJ},
  author = {The, Lih Sin and White, Simon D. M.},
  year = {1986},
  volume = {92},
  pages = {1248--1253},
  doi = {10.1086/114258}
}

@book{binney_and_tremaine_2008,
  title = {Galactic Dynamics},
  author = {Binney, James and Tremaine, Scott},
  year = {2008},
  publisher = {Princeton University Press}
}

@article{bartelmann_2010,
  title = {Gravitational lensing},
  journal = {Classical and Quantum Gravity},
  author = {Bartelmann, Matthias},
  year = {2010},
  volume = {27},
  number = {23},
  pages = {233001},
  doi = {10.1088/0264-9381/27/23/233001}
}

@inproceedings{dalton_et_al_2012,
  title = {WEAVE: The next generation wide-field spectroscopy facility for the William Herschel Telescope},
  booktitle = {SPIE Conference Series},
  author = {Gavin Dalton and Trager, \{Scott C.\} and Abrams, \{Don Carlos\} and David Carter and Piercarlo Bonifacio and Aguerri, \{J. Alfonso L.\} and Mike MacIntosh and Chris Evans and Ian Lewis and Ramon Navarro and Tibor Agocs and Kevin Dee and Sophie Rousset and Ian Tosh and Kevin Middleton and Johannes Pragt and David Terrett and Matthew Brock and Chris Benn and Marc Verheijen and \{Cano Infantes\}, Diego and Craige Bevil and Iain Steele and Chris Mottram and Stuart Bates and Gribbin, \{Francis J.\} and J{\"u}rg Rey and Rodriguez, \{Luis Fernando\} and Delgado, \{Jose Miguel\} and Isabelle Guinouard and Nic Walton and Irwin, \{Michael J.\} and Pascal Jagourel and Remko Stuik and Gerrit Gerlofsma and Ronald Roelfsma and Ian Skillen and Andy Ridings and Marc Balcells and Jean-Baptiste Daban and Carole Gouvret and Lars Venema and Paul Girard},
  year = {2012},
  doi = {10.1117/12.925950}
}

@inproceedings{Tamura_2016,
  author = {Tamura, Naoyuki and Takato, Naruhisa and Shimono, Atsushi and Moritani, Yuki and Yabe, Kiyoto and Ishizuka, Yuki and Ueda, Akitoshi and Kamata, Yukiko and others},
  title = {Prime Focus Spectrograph (PFS) for the Subaru Telescope: Overview, recent progress, and future perspectives},
  booktitle = {Ground-based and Airborne Instrumentation for Astronomy VI},
  year = {2016},
  editor = {Evans, Christopher J. and Simard, Luc and Takami, Hideki},
  series = {Proc. SPIE},
  volume = {9908},
  pages = {99081M},
  doi = {10.1117/12.2232103}
}

@article{hoekstra_et_al_2011,
  title = {Effects of distant large-scale structure on the precision of weak lensing mass measurements},
  journal = {MNRAS},
  author = {Hoekstra, Henk and Hartlap, Jan and Hilbert, Stefan and van Uitert, Edo},
  year = {2011},
  volume = {412},
  pages = {2095--2103},
  doi = {10.1111/j.1365-2966.2010.18053.x}
}

@article{rost_et_al_2021,
  title = {The Three Hundred: the structure and properties of cosmic filaments in the outskirts of galaxy clusters},
  journal = {MNRAS},
  author = {Rost, Agustin and et al.},
  year = {2021},
  volume = {502},
  pages = {714--727},
  doi = {10.1093/mnras/staa3792}
}

@article{haggar_et_al_2021,
  title = {The Three Hundred Project: Substructure in hydrodynamical and dark matter simulations of galaxy groups around clusters},
  journal = {MNRAS},
  author = {{Haggar}, Roan and {Pearce}, Frazer R. and {Gray}, Meghan E. and {Knebe}, Alexander and {Yepes}, Gustavo},
  year = {2021},
  volume = {502},
  pages = {1191--1204},
  doi = {10.1093/mnras/stab064}
}

@article{mostoghiu_et_al2021a,
  title = {The Three Hundred Project: The gas disruption of infalling objects in cluster environments},
  journal = {MNRAS},
  author = {{Mostoghiu}, Robert and {Arthur}, Jake and {Pearce}, Frazer R. and {Gray}, Meghan and {Knebe}, Alexander and {Cui}, Weiguang and {Welker}, Charlotte and {Cora}, Sof{\'\i}a A. and {Murante}, Giuseppe and {Dolag}, Klaus and {Yepes}, Gustavo},
  year = {2021},
  volume = {501},
  pages = {5029--5041},
  doi = {10.1093/mnras/stab014}
}

@article{capalbo_et_al_2025,
  title = {Inference of the morphology and dynamical state of nearby Planck-SZ galaxy clusters with Zernike polynomials},
  journal = {\aap},
  author = {{Capalbo}, V. and {De Petris}, M. and {Ferragamo}, A. and {Cui}, W. and {Ruppin}, F. and {Yepes}, G.},
  year = {2025},
  volume = {698},
  pages = {A201},
  doi = {10.1051/0004-6361/202452649}
}

@article{sayers_et_al_2021,
  title = {CLUMP-3D: the lack of non-thermal motions in galaxy cluster cores},
  journal = {MNRAS},
  author = {{Sayers}, Jack and {Sereno}, Mauro and {Ettori}, Stefano and {Rasia}, Elena and {Cui}, Weiguang and {Golwala}, Sunil and {Umetsu}, Keiichi and {Yepes}, Gustavo},
  year = {2021},
  volume = {505},
  pages = {4338--4344},
  doi = {10.1093/mnras/stab1542}
}

@article{sereno_et_al_2021,
  title = {The thermalization of massive galaxy clusters},
  journal = {MNRAS},
  author = {{Sereno}, Mauro and {Lovisari}, Lorenzo and {Cui}, Weiguang and {Schellenberger}, Gerrit},
  year = {2021},
  volume = {507},
  pages = {5214--5223},
  doi = {10.1093/mnras/stab2435}
}

@article{li_et_al_2022,
       author = {Li, Qingyang and Han, Jiaxin and Wang, Wenting and Cui, Weiguang and De Luca, Federico and Yang, Xiaohu and Zhou, Yanrui and Shi, Rui},
        title = {What to expect from dynamical modelling of cluster haloes -- II. Investigating dynamical state indicators with Random Forest},
      journal = {MNRAS},
         year = {2022},
       volume = {514},
       number = {4},
        pages = {5890-5904},
          doi = {10.1093/mnras/stac1739}
}

@article{nelson_2024,
  title = {Introducing the TNG-Cluster simulation: Overview and first results},
  journal = {A\&A},
  author = {{Nelson}, Dylan and {Pillepich}, Annalisa and {Ayromlou}, Mohammadreza and {Lee}, Wonki and {Lehle}, Katrin and {Rohr}, Eric and {Truong}, Nhut},
  year = {2024},
  volume = {670},
  pages = {A1},
  doi = {10.1051/0004-6361/202348623}
}

@article{silverman_1986,
  title = {Resolution of a cosmological paradox using general relativity},
  journal = {American Journal of Physics},
  author = {Silverman, Abraham N.},
  year = {1986},
  volume = {54},
  pages = {1092--1096},
  doi = {10.1119/1.14721}
}

@article{sifon_chances_2025,
  title = {CHANCES: The Chilean Cluster Galaxy Evolution Survey},
  journal = {A\&A},
  author = {Sifón, Cristóbal and Finoguenov, Alexis and Haines, Christopher P. and Jaffé, Yara and Amrutha, B. M. and Demarco, Ricardo and Lima, E. V. R. and Lima-Dias, Ciria and Méndez-Hernández, Hugo and Merluzzi, Paola and Monachesi, Antonela and Teixeira, Gabriel S. M. and Tejos, Nicolas and Almeida-Fernandes, F. and Araya-Araya, Pablo and Argudo-Fernández, Maria and Baier-Soto, Raúl and Bilton, Lawrence E. and Bom, C. R. and Calderón, Juan Pablo and Cassarà, Letizia P. and Comparat, Johan and Courtois, H. M. and D’Ago, Giuseppe and Dupuy, Alexandra and Fritz, Alexander and Haack, Rodrigo F. and Herpich, Fabio R. and Ibar, E. and Kuchner, Ulrike and Lacerna, Ivan and Lopes, Amanda R. and Lopez, Sebastian and Lösch, Elismar and McGee, Sean and Mendes De Oliveira, C. and Morelli, Lorenzo and Moretti, Alessia and Pallero, Diego and Piraino-Cerda, Franco and Pompei, Emanuela and Rescigno, U. and Smith Castelli, Analía V. and Smith, Rory and Sodré Jr, Laerte and Tempel, Elmo},
  year = {2025},
  volume = {697},
  pages = {A92},
  doi = {10.1051/0004-6361/202452710}
}

@INPROCEEDINGS{Armando_2024,
       author = {{Gil de Paz}, A. and {Iglesias-P{\'a}ramo}, J. and {Carrasco}, E. and {Gallego}, J. and {Garc{\'\i}a Vargas}, M.~L. and {Hern{\'a}ndez}, L. and {Mart{\'\i}n Garz{\'o}n}, G.~E. and {O{\~n}orbe}, J. and {Piqueras L{\'o}pez}, J. and {V{\'\i}lchez}, J.~M. and {S{\'a}nchez-Bl{\'a}zquez}, P. and {Kehrig}, C. and {Monta{\~n}a}, A. and {Montenegro Montes}, F.~M. and {P{\'e}rez-Calpena}, A. and {Tulloch}, S. and {Abril}, M. and {Pascual}, S. and {Cardiel}, N. and {Castillo-Morales}, {\'A}. and {Jim{\'e}nez-Teja}, Y. and {Gonz{\'a}lez Ru{\'\i}z}, V. and {P{\'e}rez Medialdea}, D. and {Calvo-Ortega}, R. and {P{\'a}ez}, G.},
        title = "{TARSIS, the 8 arcmin$^{2}$ IFU for the Calar Alto 3.5m telescope}",
    booktitle = {Ground-based and Airborne Instrumentation for Astronomy X},
         year = 2024,
       editor = {{Bryant}, Julia J. and {Motohara}, Kentaro and {Vernet}, Jo{\"e}l. R.~D.},
       series = {Society of Photo-Optical Instrumentation Engineers (SPIE) Conference Series},
       volume = {13096},
        month = jul,
        pages = {1309620},
          doi = {10.1117/12.3016123}
}

@article{MamonLokas_2003,
  title = {Dark matter distribution in the Coma cluster from galaxy kinematics},
  journal = {MNRAS},
  author = {Łokas, Ewa L. and Mamon, Gary A.},
  year = {2003},
  volume = {343},
  pages = {401--412},
  doi = {10.1046/j.1365-8711.2003.06684.x}
}

@article{Lokas2006,
  title = {Mass distribution in nearby Abell clusters},
  journal = {MNRAS},
  author = {{{\L}okas}, E.~L. and {Wojtak}, R. and {Gottl{\"o}ber}, S. and {Mamon}, G.~A. and {Prada}, F.},
  year = {2006},
  volume = {367},
  pages = {1463--1472},
  doi = {10.1111/j.1365-2966.2006.10151.x}
}

@article{MamonBoue2010,
  title = {Kinematic deprojection and mass inversion of spherical systems},
  journal = {MNRAS},
  author = {Mamon, Gary A. and Boué, Gwenaël},
  year = {2010},
  volume = {401},
  pages = {2433--2450},
  doi = {10.1111/j.1365-2966.2009.15817.x}
}

@article{Hopkins2015,
  title = {A new class of accurate mesh-free hydrodynamic simulation methods},
  journal = {MNRAS},
  author = {Hopkins, Philip F.},
  year = {2015},
  volume = {450},
  pages = {53--110},
  doi = {10.1093/mnras/stv195}
}

@article{beck2016,
  title = {An improved SPH scheme for cosmological simulations},
  journal = {MNRAS},
  author = {{Beck}, A.~M. and {Murante}, G. and {Arth}, A. and {Remus}, R. -S. and {Teklu}, A.~F. and {Donnert}, J.~M.~F. and {Planelles}, S. and {Beck}, M.~C. and {F{\"o}rster}, P. and {Imgrund}, M. and {Dolag}, K. and {Borgani}, S.},
  year = {2016},
  volume = {455},
  pages = {2110--2130},
  doi = {10.1093/mnras/stv2443}
}

@article{deAndres_2022,
  author = {{de Andres}, Daniel and {Cui}, Weiguang and {Ruppin}, Florian and {De Petris}, Marco and {Yepes}, Gustavo and {Gianfagna}, Giulia and {Lahouli}, Ichraf and {Aversano}, Gianmarco and {Dupuis}, Romain and {Jarraya}, Mahmoud and {Vega-Ferrero}, Jesús},
  title = {A deep learning approach to infer galaxy cluster masses from Planck Compton-y parameter maps},
  journal = {Nature Astronomy},
  year = {2022},
  volume = {6},
  pages = {1325--1331},
  doi = {10.1038/s41550-022-01784-y}
}

@article{Gianfagna_2023,
  author = {{Gianfagna}, Giulia and {Rasia}, Elena and {Cui}, Weiguang and {De Petris}, Marco and {Yepes}, Gustavo and {Contreras-Santos}, Ana and {Knebe}, Alexander},
  title = {A study of the hydrostatic mass bias dependence and evolution within The Three Hundred clusters},
  journal = {MNRAS},
  year = {2023},
  volume = {518},
  number = {3},
  pages = {4238--4248},
  doi = {10.1093/mnras/stac3364}
}

@article{Ferragamo_2023,
  author = {{Ferragamo}, A. and {de Andres}, D. and {Sbriglio}, A. and {Cui}, W. and {De Petris}, M. and {Yepes}, G. and {Dupuis}, R. and {Jarraya}, M. and {Lahouli}, I. and {De Luca}, F. and {Gianfagna}, G. and {Rasia}, E.},
  title = {THE THREE HUNDRED project: a machine learning method to infer clusters of galaxy mass radial profiles from mock Sunyaev-Zel'dovich maps},
  journal = {MNRAS},
  year = {2023},
  volume = {520},
  number = {3},
  pages = {4000--4008},
  doi = {10.1093/mnras/stad377}
}

@article{Iqbal_2025,
  author = {{Iqbal}, Asif and {Majumdar}, Subhabrata and {Rasia}, Elena and {Pratt}, Gabriel W. and {de Andres}, Daniel and {Melin}, Jean-Baptiste and {Cui}, Weiguang},
  title = {Deriving accurate galaxy cluster masses using X-ray thermodynamic profiles and graph neural networks},
  journal = {A\&A},
  year = {2025},
  volume = {704},
  pages = {A334},
  doi = {10.1051/0004-6361/202554321}
}

@article{cui_halo_2016,
  author = {Cui, Weiguang and Knebe, Alexander and Yepes, Gustavo and others},
  title = {The halo boundary of virialized structures},
  journal = {MNRAS},
  year = {2016},
  volume = {456},
  pages = {2566--2582},
  doi = {10.1093/mnras/stv2807}
}

@article{kennicutt_1998,
  author = {Kennicutt, R. C.},
  title = {Star Formation in Galaxies Along the Hubble Sequence},
  journal = {ARA\&A},
  year = {1998},
  volume = {36},
  pages = {189--232},
  doi = {10.1146/annurev.astro.36.1.189}
}

@article{yoon_outskirts_2022,
  author = {Yoon, J. H. and Rasia, E. and Biffi, V. and Dolag, K. and Planelles, S. and Munari, E. and Fabjan, D. and Beck, A. M. and Dalla Vecchia, C. and Ragone-Figueroa, C.},
  title = {The outskirts of galaxy clusters with MUSIC simulations: pressure profiles, clumping and hydrostatic bias},
  journal = {MNRAS},
  year = {2022},
  volume = {516},
  number = {3},
  pages = {4084--4100},
  doi = {10.1093/mnras/stac2432}
}

@article{Andreon_2016,
  author = {Andreon, S.},
  title = {Richness-based masses of rich and famous galaxy clusters},
  journal = {A\&A},
  year = {2016},
  volume = {587},
  pages = {A158},
  doi = {10.1051/0004-6361/201526852}
}

@article{Li_et_al_2021,
  author = {Li, Qingyang and Han, Jiaxin and Wang, Wenting and Cui, Weiguang and Li, Zhaozhou and Yang, Xiaohu},
  title = {What to expect from dynamical modelling of cluster haloes -- I. The information content of different dynamical tracers},
  journal = {MNRAS},
  year = {2021},
  volume = {505},
  number = {3},
  pages = {3907-3922},
  doi = {10.1093/mnras/stab1633}

}

@article{Contreras_2024,
  author = {Contreras-Santos, A. and Knebe, A. and Cui, W. and Alonso Asensio, I. and Dalla Vecchia, C. and Ca{\~n}as, R. and Haggar, R. and Mostoghiu Paun, R. A. and Pearce, F. R. and Rasia, E.},
  title = {Characterising the intra-cluster light in The Three Hundred simulations},
  journal = {A\&A},
  year = {2024},
  volume = {683},
  pages = {A59},
  doi = {10.1051/0004-6361/202348474}
}

@article{DeAndres_2024,
  author = {de Andres, Daniel and Cui, Weiguang and Yepes, Gustavo and De Petris, Marco and Ferragamo, Antonio and De Luca, Federico and Aversano, Gianmarco and Rennehan, Douglas},
  title = {The three hundred project: mapping the matter distribution in galaxy clusters via deep learning from multiview simulated observations},
  journal = {MNRAS},
  year = {2024},
  volume = {528},
  number = {2},
  pages = {1517--1530},
  doi = {10.1093/mnras/stae071}
}

@article{KnollmannKnebe_2009,
  author = {Knollmann, Steffen R. and Knebe, Alexander},
  title = {AHF: Amiga's Halo Finder},
  journal = {ApJ Supplement Series},
  year = {2009},
  volume = {182},
  number = {2},
  pages = {608--624},
  doi = {10.1088/0067-0049/182/2/608}
}

@article{munari_effects_2013,
       author = {{Munari}, E. and {Biviano}, A. and {Borgani}, S. and {Murante}, G. and {Fabjan}, D.},
        title = "{The relation between velocity dispersion and mass in simulated clusters of galaxies: dependence on the tracer and the baryonic physics}",
      journal = {MNRAS},
         year = 2013,
       volume = {430},
       number = {4},
        pages = {2638-2649},
          doi = {10.1093/mnras/stt049}
}

@article{serra_identification_2013,
       author = {{Serra}, Ana Laura and {Diaferio}, Antonaldo},
        title = "{Identification of Members in the Central and Outer Regions of Galaxy Clusters}",
      journal = {\apj},
         year = 2013,
       volume = {768},
       number = {2},
          eid = {116},
        pages = {116},
          doi = {10.1088/0004-637X/768/2/116}
}

@article{Li_2021,
       author = {{Li}, Qingyang and {Han}, Jiaxin and {Wang}, Wenting and {Cui}, Weiguang and {Li}, Zhaozhou and {Yang}, Xiaohu},
        title = "{What to expect from dynamical modelling of cluster haloes - I. The information content of different dynamical tracers}",
      journal = {MNRAS},
         year = 2021,
        month = aug,
       volume = {505},
       number = {3},
        pages = {3907-3922},
          doi = {10.1093/mnras/stab1633}
}

\end{document}